\documentclass[aps,preprint,onecolumn,secnumarabic,nobalancelastpage,amsmath,amssymb,
nofootinbib]{revtex4}

\RequirePackage{fix-cm}
\usepackage{enumerate}

\usepackage{pst-plot}
\usepackage[scriptsize,nooneline,hang]{caption}
\usepackage[hang,nooneline,scriptsize]{subfigure}
\usepackage{graphics}      % standard graphics specifications
\usepackage{graphicx}      % alternative graphics specifications
\usepackage{url}           % for on-line citations
\usepackage{dcolumn}% Align table columns on decimal point
\usepackage{hyperref} %add hypertext capabilities
\usepackage{bm}            % special 'bold-math' package
\usepackage{mathrsfs}
\usepackage{epstopdf}
\usepackage{color}
\begin{document}

reprint{APS/123-QED}

\title{Geodesic equations in the static and rotating dilaton black holes: Analytical solutions and applications }

\author{Saheb Soroushfar}
\author{Reza Saffari}%
\email{rsk@guilan.ac.ir}
\author{Ehsan Sahami}
\affiliation{Department of Physics, University of Guilan,
41335-1914, Rasht, Iran.
%${}^2$Institut f\"ur Physik, Universit\"at Oldenburg, Postfach 2503 D-26111 Oldenburg, Germany.
}%


%\author{Jutta Kunz}
%\affiliation{
 %Department of Physics, University of Guilan, P. O. Box: 41335-1914, Rasht, Iran.
 %\\
%\affiliation{
 %Third institution, the second for Charlie Author
%\affiliation{%

%\author{Saheb Soroushfar\altaffilmark{1}}
% \altaffiliation{soroush@mail.yu.ac.ir}%Lines break automatically or can be forced with \\
%\author{Reza Saffari\altaffilmark{1}}%
%\and
%\author{Jutta Kunz\altaffilmark{2}}
%\email{rsk@guilan.ac.ir}
%\author{Jutta Kunz\altaffilmark{2}}%
%\altaffiltext{1}{%
 %Department of Physics, University of Guilan, P. O. Box: 41335-1914, Rasht, Iran.
%\altaffiltext{2}{aaa.}
%\author         {Saheb Soroushfar\dag, Reza Saffari}
%\email          {soroush@mail.yu.ac.ir}
%\email       {rsk@guilan.ac.ir}
%\affiliation    {Department of Physics, University of Guilan, P. O. Box: 41335-1914, Rasht, Iran\dag,
%Physics Department, College of Sciences, Yasouj University, 75914 Yasouj, Iran }
\date{\today}

\begin{abstract}

In this paper, we consider the timelike and null geodesics around 
the static [GMGHS (Gibbons, Maeda, Garfinkle, Horowitz and Strominger), 
magnetically charged GMGHS, electrically charged GMGHS]  
and the rotating (Kerr-Sen dilaton-axion) dilaton black holes.  
The geodesic equations are solved in terms of Weierstrass elliptic functions. 
To classify the trajectories around the black holes, we use the
analytical solution and effective potential techniques and then characterize the 
different types of the resulting orbits in terms of the conserved energy and angular
momentum. Also, using the obtained results we study astrophysical applications.
\end{abstract}

\maketitle

%%%%%%%%%%%%%%%%%%%%%%%%%%%%%%%%%%%%%%%%%%%%%%%%%%%%%%%%%%%%%%%%%%%%%%%%%%%%%
\section{INTRODUCTION}

The well-known exact solution of the vacuum Einstein equations
described by Schwarzschild in 1916 \cite{K.Schwarzschild} as a
spherically symmetric black hole in a four dimensional spacetime.
Addition of an electric charge, change the Schwarzschild solution to
a charged black hole. This solution was discovered by Reissner
(1916), Weyl (1917) and Nordstr\"om (1918), independently and now it
is known as Reissner-Nordstr\"om metric \cite{Blaga:2014spa}. Also,
another solution of charged black hole in four dimensions was
obtained by Gibbons and Maeda \cite{Gibbons:1987ps}, independently,
by Garfinkle, Horowitz and Strominger \cite{Garfinkle:1990qj} using
a scalar field in range of low-energy of heterotic string theory,
which is called GMGHS solution \cite{Mukherjee&Majumdar}. The GMGHS
black hole can be explained in string or Einstein frame, which are
connected to each other by conformal transformation despite of
differences of the physical properties in each frames
\cite{Faraoni&Gunzig&Nardone,Casadio and Harms,Kim&Choi&Park}.

Study of motion of massive and massless particles give a set of
comprehensive information about the gravitational field around a
black hole. Analysis of geodesic equation of motion predict some
observational phenomena such as perihelion shift, gravitational
time-delay and light deflection. The first analytic solution for
Schwarzschild spacetime using Weierstrassian elliptic functions and
their derivatives presented by Hagihara in 1931 \cite{Y. Hagihara}.
The theoretical and mathematical properties of Weierstrassian
elliptic functions demonstrated by Jacobi \cite{Jacobi:1841kw}, Abel
\cite{N.H. Abel}, Riemann \cite {B. Riemann1857,B. Riemann1866},
Weierstrass \cite{K.T.W. Weierstrass} and Baker \cite{H.F. Baker}.

Analytical solutions of geodesic equations were investigated for
different spacetimes such as Reissner-Nordstr\"om,
Schwarzschild-(anti)de Sitter and Reissner-Nordstr\"om--(anti)--de
Sitter spactime in four dimensions and in higherdimensions
\cite{Chandrasekhar:1985kt,Hackmann:2008zz,Hackmann:2008tu}. Also
the motion of test particles around rotating black holes
\cite{Hackmann:2010zz,Kagramanova:2010bk} and in the spacetime of a
black hole which is combined by cosmic string was studied
extensively \cite{Hackmann:2009rp,Hackmann:2010ir}. Recently,
geodesic equations were solved analytically in the spacetime of
black hole in f(R) gravity
\cite{Soroushfar:2015wqa,Soroushfar:2015dfz}. 
Analysis of geodesics, include null, timelike
\cite{Fernando:2011ki,Olivares:2013jza}, circular null and timelike
geodesics \cite{Pradhan:2012id,Blaga:2014lva}, were studied in the
spacetime of GMGHS black hole in the special cases.

The aim of this paper is to determine the complete set of analytic
solutions of the geodesic equations in the spacetime of the static (GMGHS,
magnetically charged GMGHS, electrically charged GMGHS)  and rotating 
(Ker-Sen Dilaton-Axion) dilaton black holes. 
We discussed the motion of test particles and light rays in the
spacetime of these black holes and present the analytic solutions of
the geodesic equations in terms of the elliptic Weierstrass
functions. Then we determined the type of the orbits for test 
particles and light rays in the vicinity of these static and rotating dilaton black holes. 
In the first part the static dilaton black holes is studied and In the 
second part the rotating dilaton black hole is analysed. 

Our paper is organized as follows: In Sec.~(\ref{Static Dilaton}), we
introduce the metrics and their histories for static dilaton black holes. 
Then we derive the geodesic equations from Lagrangian corresponding to the metric and discuss the
effective potentials. We solve  geodesic equations and classify the
solutions of timelike and null geodesic equations. Using analytical solutions and 
effective potentials, we plot some possible orbits for test particles 
around each black hole in acceptable regions. At the end of 
this section, we study Astrophysical applications. 
 In Sec.~(\ref{Rotating Dilaton}), we introduce the metric 
for a rotating dilaton black hole. Then we derive the geodesic equations 
and effective potential. We solve the geodesic equations analytically and 
plot the possible orbits. Our conclusions are drawn in Sec.~(\ref{conclusions}).

%%%%%%%%%%%%%%%%%%%%%%%%%%%
% Next section
%%%%%%%%%%%%%%%%%%%%%%%%%%%

\section{Static Dilaton black holes}\label{Static Dilaton}

In this section, we will discuss the geodesics in the static dilaton black holes and present analytical
solutions of the equations of motion.

\subsection{Metrics}\label{metrics}

In this section, we review all the spacetimes which are used as static dilaton black holes. 
In the Einstein frame, the GMGHS action is
\cite{Garfinkle:1990qj}
\begin{equation}\label{S-Ein}
S=\int d^{4}x\sqrt{-g}(R-2(\bigtriangledown\phi)^{2}-e^{-2\phi}F_{\mu\nu}F^{\mu\nu}),
\end{equation}
where $\phi$ is a dilaton, $R$ is the scalar curvature, and
$F_{\mu\nu}$ is the Maxwell field. The spherically symmetric static
charged solutions to equations of motion of the action (\ref{S-Ein})
is 
\begin{equation}\label{metric GM}
ds_{GM}^{2}=-(1-\frac{2M}{r})dt^{2}+(1-\frac{2M}{r})^{-1}dr^{2}+r(r-\frac{Q^{2}}{M})(d\theta^{2}+\sin^{2}\theta d\varphi^{2}) ,
\end{equation}
where $M$ and $Q$ are mass and charge, respectively. This solution
was obtained by Gibbons and Maeda, and independently Garfinkle,
Horowitz and Strominger with use a transformation to the
Schwarzschild solution \cite{Harrison,Blaga:2014spa}.

The GMGHS action in the string frame, is 
\begin{align}
S=\int d^{4}x\sqrt{-g}e^{-2\phi}(R+4(\bigtriangledown\phi)^{2}-F_{\mu \nu}F^{\mu \nu}) ,
\end{align}
where $\phi$ is a dilaton, R is the scalar curvature, and
$F_{\mu\nu}$ is the Maxwell's field strength. String frame is
related to the Einstein frame action by the conformal transformation
of $g^{s}_{\mu\nu}=e^{2\phi}g^{E}_{\mu\nu}$,
\cite{Gibbons:1987ps,Garfinkle:1990qj}. By going from an
electrically to a magnetically charged black hole, the string metric
does change with the change in sign of dilaton $\phi$, but the
Einstein metric does not change. Thus, the magnetically charged
GMGHS black hole metric in the string frame is given by
\cite{Kim&Choi&Park,Choi:2014wna}:
\begin{align}\label{metric Mag}
ds_{Mag}^{2}=-\dfrac{(1-\dfrac{2M}{r})}{(1-\dfrac{Q^{2}}{Mr})}
dt^{2}+\dfrac{dr^{2}}{(1-\dfrac{2M}{r})(1-\dfrac{Q^{2}}{Mr})}+r^{2}(d\theta
^{2}+sin^{2}\theta d\varphi ^{2}) ,
\end{align}
And the electrically charged GMGHS solution in the string frame is
given by:
\begin{align}\label{metric El}
ds_{El}^{2}=-\dfrac{(1+\dfrac{Q^2-2M^2}{Mr})}{(1+\dfrac{Q^{2}}{Mr})^{2}}dt^{2}+\dfrac{dr^{2}}{(1+\dfrac{Q^{2}-2M^2}{Mr})}+r^2(d\theta
^{2}+sin^2\theta d\varphi^{2}).
\end{align}

\subsection{The geodesic equations}\label{geodesic}

The geodesic equations can be derived by compute the Lagrangian for each
metric as \cite{Chandrasekhar:1985kt}
\begin{equation}
\mathcal{L}=\dfrac{1}{2}
g_{\mu\nu}\dfrac{dx^{\mu}}{ds}\dfrac{dx^{\nu}}{ds}=
\dfrac{1}{2}\epsilon ,
\end{equation}
where $ \epsilon=0,1 $ for null and timelike geodesics respectively.
Thus the Lagrangian for the metric (\ref{metric GM}) is:
\begin{align}\label{lagrangy-GM}
2\mathcal{L}_{GM}=-(1-\dfrac{2M}{r})
\dot{t}^{2}+(1-\dfrac{2M}{r})^{-1} \dot{r}^{2}+r(r-\dfrac{Q^{2}}{M})
\dot{\theta}^{2}+r(r-\dfrac{Q^{2}}{M})sin^{2}\theta
\dot{\varphi}^{2} ,
\end{align}
for the metric (\ref{metric Mag}) is:
\begin{align}\label{lagrangy-Mag}
2\mathcal{L}_{Mag}=-\dfrac{(1-\dfrac{2M}{r})}{(1-\dfrac{Q^{2}}{Mr})}
\dot{t}^{2}+\dfrac{1}{(1-\dfrac{2M}{r})(1-\dfrac{Q^{2}}{Mr})}\dot{r}^{2}+r^{2}(\dot{\theta}
^{2}+sin^{2}\theta \dot{\varphi} ^{2}) ,
\end{align}
and for the metric (\ref{metric El}) is:
\begin{align}\label{lagrangy-El}
2\mathcal{L}_{El}=-\dfrac{(1+\dfrac{Q^2-2M^2}{Mr})}{(1+\dfrac{Q^{2}}{Mr})^{2}}\dot{t}^{2}+\dfrac{1}{(1+\dfrac{Q^{2}-2M^2}{Mr})}
\dot{r}^{2}+r^2(\dot{\theta} ^{2}+sin^2\theta \dot{\varphi}^{2}).
\end{align}
The Killing vectors respect to the spacetime from the Euler-Lagrange
for time and latitude are $\dfrac{\partial}{\partial t}$ and
$\dfrac{\partial}{\partial \varphi}$. The energy $E$ and the angular
momentum $L$ are the constants of motion which are given by the
generalized momenta $P_{t}$ and $P_{\varphi}$
\begin{equation}
P_{t}=\dfrac{\partial \mathcal{L}}{\partial\dot{t}}=-E,  \qquad
P_{\varphi}=\dfrac{\partial \mathcal{L}}{\partial\dot{\varphi}}=L.
\end{equation}
From the Euler-Lagrange equation for $t$ we get to the energy
conservations
\begin{equation}\label{dt/ds- GM}
E_{GM}=(g_{tt})_{GM}\dfrac{dt}{ds}=-(1-\dfrac{2M}{r}) \dfrac{dt}{ds} ,
\end{equation}
\begin{equation}\label{dt/ds- Mag}
E_{Mag}=(g_{tt})_{Mag}\dfrac{dt}{ds}=
-\dfrac{(1-\dfrac{2M}{r})}{(1-\dfrac{Q^{2}}{Mr})}\dfrac{dt}{ds} ,
\end{equation}
\begin{equation}\label{dt/ds- El}
E_{El}=(g_{tt})_{El}\dfrac{dt}{ds}=-\dfrac{(1+\dfrac{Q^2-2M^2}{Mr})}
{(1+\dfrac{Q^{2}}{Mr})^{2}}\dfrac{dt}{ds} ,
\end{equation}
and for $\varphi$ we obtained the angular momentum conservations
\begin{equation}\label{dfi/ds- GM}
L_{GM}=(g_{\varphi\varphi})_{GM}\dfrac{d\varphi}{ds}=
r(r-\dfrac{Q^{2}}{M})sin^{2}\theta \dfrac{d\varphi}{ds} ,
\end{equation}
\begin{equation}\label{dfi/ds- Mag}
L_{Mag}=(g_{\varphi\varphi})_{Mag}\dfrac{d\varphi}{ds}= r^{2}
sin^{2}\theta \dfrac{d\varphi}{ds} ,
\end{equation}
\begin{equation}\label{dfi/ds- El}
L_{El}=(g_{\varphi\varphi})_{El}\dfrac{d\varphi}{ds}= r^{2}
sin^{2}\theta \dfrac{d\varphi}{ds}.
\end{equation}
We consider the motion is took place in a equatorial plane because
of the existence of spherically symmetry and choose
$\theta=\frac{\pi}{2}$ and $\dot{\theta}=0$ as the initial
conditions. Therefore with substitute $\dot{t}$ and $\dot{\varphi}$
from Eqs.(\ref{dt/ds- GM})-(\ref{dfi/ds- El}), in
Eqs.(\ref{lagrangy-GM})-(\ref{lagrangy-El}), we get
\begin{equation}\label{dr/ds- GM}
(\dfrac{dr}{ds})_{GM}^{2}=E_{GM}^{2}-(1-\dfrac{2M}{r})
(\dfrac{L_{GM}^{2}}{r(r-\frac{Q^{2}}{M})}+\epsilon ) ,
\end{equation}
\begin{equation}\label{dr/ds- Mag}
(\dfrac{dr}{ds})_{Mag}^{2}=(1-\dfrac{Q^{2}}{Mr})^{2} E_{Mag}^{2} -
(1-\dfrac{2M}{r})(1-\dfrac{Q^{2}}{Mr})
(\dfrac{L_{Mg}^{2}}{r^{2}}+\epsilon) ,
\end{equation}
\begin{equation}\label{dr/ds- El}
(\dfrac{dr}{ds})_{El}^{2}=(1+\dfrac{Q^{2}}{Mr})^{2}E_{El}^{2}
-(1+\dfrac{Q^{2}-2M^{2}}{Mr})(\dfrac{L_{El}^{2}}{r^{2}}+\epsilon).
\end{equation}
We obtain the corresponding equation for $r$ as a function of
$\varphi$ and as a function of $t$, with energy and angular momentum
conservation
\begin{align}\label{dr/dfi- GM}
(\dfrac{dr}{d\varphi})^{2}_{GM}=(\dfrac{E^{2}}{L^{2}}
-\dfrac{\epsilon}{L^{2}})r^{4}+(\dfrac{-2E^{2}Q^{2}}{ML^{2}}+\dfrac{2\epsilon
Q^{2}}{ML^{2}}+\dfrac{2M\epsilon}{L^{2}})r^{3}+
(\dfrac{E^{2}Q^{4}}{M^{2}L^{2}}-\nonumber\\-\dfrac{\epsilon
Q^{4}}{M^{2}L^{2}}-\dfrac{4\epsilon
Q^{2}}{L^{2}}-1)r^{2}+(\dfrac{2\epsilon
Q^{4}}{ML^{2}}+\dfrac{Q^{2}}{M}+2M)r-2 Q^{2}=R_{GM}(r) ,
\end{align}
\begin{align}\label{dr/dfi- Mag}
(\dfrac{dr}{d\varphi})^{2}_{Mag}=(\dfrac{E^{2}}{L^{2}}
-\dfrac{\epsilon}{L^{2}})r^{4}+(\dfrac{-2E^{2}Q^{2}}{ML^{2}}+\dfrac{\epsilon
Q^{2}}{ML^{2}}+\dfrac{2M\epsilon}{L^{2}})r^{3}+\nonumber\\
+(\dfrac{E^{2}Q^{4}}{M^{2}L^{2}}-\dfrac{2\epsilon
Q^{2}}{L^{2}}-1)r^{2}+(\dfrac{Q^{2}}{M}+2M)r-2 Q^{2}=R_{Mag}(r) ,
\end{align}
\begin{align}\label{dr/dfi- El}
(\dfrac{dr}{d\varphi})^{2}_{El}=(\dfrac{E^{2}}{L^{2}}
-\dfrac{\epsilon}{L^{2}})r^{4}+(\dfrac{2E^{2}Q^{2}}{ML^{2}}-\dfrac{\epsilon
Q^{2}}{ML^{2}}+\dfrac{2M\epsilon}{L^{2}})r^{3}+\nonumber\\
+(\dfrac{E^{2}Q^{4}}{M^{2}L^{2}}-1)r^{2}+(-\dfrac{Q^{2}}{M}+2M)r=
R_{El}(r) ,
\end{align}
and
\begin{align}\label{dr/dt- GM}
(\dfrac{dr}{dt})_{GM}^{2}=\frac{1}{E^{2}}(1-\frac{2M}{r})^{2}[E^{2}
-(1-\frac{2M}{r})(\dfrac{L^{2}}{r(r-\frac{Q^{2}}{M})}+\epsilon)],
\end{align}
\begin{align}\label{dr/dt- Mag}
(\dfrac{dr}{dt})_{Mag}^{2}=\frac{1}{E^{2}}\dfrac{(1-\frac{2M}{r})^{2}}
{(1-\dfrac{Q^{2}}{Mr})^{2}}[(1-\dfrac{Q^{2}}{Mr})^{2}
E^{2}-(1-\frac{2M}{r})(1-\frac{Q^{2}}{Mr})(\frac{L^{2}}{r^{2}}+\epsilon)],
\end{align}
\begin{align}\label{dr/dt- El}
(\dfrac{dr}{dt})_{El}^{2}=\frac{1}{E^{2}}
\dfrac{(1+\dfrac{Q^{2}-2M^{2}}{Mr})^{2}}{(1+\frac{Q^{2}}{Mr})^{2}}
[(1+\frac{Q^{2}}{Mr})^{2}E^{2}-(1+\dfrac{Q^{2}-2M^{2}}{Mr})(\frac{L^{2}}{r^{2}}+\epsilon)].
\end{align}
Eqs.(\ref{dr/ds- GM})-(\ref{dr/dt- El}) gives a complete definition
of the dynamics. In these set of equations, the values of $L$ and $E^{2}$ in the right hand side 
refer to indices of the left hand side of them, that we ignore indices for $L$ and $E^{2}$  for simplicity.
We get the effective potential by comparing
Eqs.(\ref{dr/ds- GM})-(\ref{dr/ds- El}) with 
$\dot{r}^{2} + V_{eff}= E^{2}$,
\begin{equation}
(V_{eff})_{GM}=(1-\dfrac{2M}{r})(\dfrac{L^{2}}{r(r-\frac{Q^{2}}{M})}+\epsilon ) ,
\end{equation}
\begin{equation}
(V_{eff})_{Mag}=\dfrac{(1-\dfrac{2M}{r})
(\dfrac{L^{2}}{r^{2}}+\epsilon)}{(1-\dfrac{Q^{2}}{Mr})} ,
\end{equation}
\begin{equation}
(V_{eff})_{El}=-\dfrac{(1+\dfrac{Q^2-2M^2}{Mr})
(\dfrac{L^{2}}{r^{2}}+\epsilon)}{(1+\dfrac{Q^{2}}{Mr})^{2}} ,
\end{equation}

which depends on radial coordinate
 $r$, charge $Q$ and mass $M$ of
the black hole, the type of the geodesics $\epsilon$ and the angular
momentum $L$ of the particles.  In these set of equations, again the value of $L$  in the right hand side 
refer to indices of the left hand side of them, that we ignore indices for $L$ and $E^{2}$  for simplicity.

We introduce dimensionless quantities for rescale the parameters
\begin{equation}
\tilde{r}=\frac{r}{M}  , \qquad \tilde{Q} = \frac{Q}{M}  ,\qquad
\tilde{L}=\dfrac{M^{2}}{L^{2}},
\end{equation}
and rewrite Eqs.(\ref{dr/dfi- GM})-(\ref{dr/dfi- El}) as
\begin{align}\label{drt/dfi- GM}
(\dfrac{d \tilde{r}}{d\varphi})^{2}_{GM}=(E^{2}-\epsilon)\tilde{L}
\tilde{r}^{4}+(2 \tilde{Q}^{2}\epsilon + 2\epsilon -2 E^{2}
\tilde{Q}^{2}) \tilde{L} \tilde{r}^{3} + (E^{2} \tilde{Q}^{4}
\tilde{L} \nonumber\\-\epsilon \tilde{Q}^{4}\tilde{L} -4
\tilde{Q}^{2} \epsilon \tilde{L} -1) \tilde{r}^{2} +(2\epsilon
\tilde{Q}^{4} \tilde{L} +\tilde{Q}^{2} +2) \tilde{r} -2
\tilde{Q}^{2} =R_{GM}(\tilde{r}) ,
\end{align}
\begin{align}\label{drt/dfi- Mag}
(\dfrac{d \tilde{r}}{d\varphi})^{2}_{Mag}=(E^{2} -\epsilon)
\tilde{L} \tilde{r}^{4} +(\tilde{Q}^{2} \epsilon -2 \tilde{Q}^{2}
E^{2} +2\epsilon)\tilde{L} \tilde{r}^{3} +\nonumber\\ (\tilde{Q}^{4}
E^{2} \tilde{L} - 2\tilde{Q}^{2}\epsilon
\tilde{L}-1)\tilde{r}^{2}+(\tilde{Q}^{2} +2)\tilde{r}
-2\tilde{Q}^{2} =R_{Mag}(\tilde{r}) ,
\end{align}
\begin{align}\label{drt/dfi- El}
(\dfrac{d \tilde{r}}{d\varphi})^{2}_{El}=(E^{2} -\epsilon) \tilde{L}
\tilde{r}^{4} +(2 \tilde{Q}^{2} E^{2}-\tilde{Q}^{2} \epsilon
+2\epsilon)\tilde{L} \tilde{r}^{3} \nonumber\\+(\tilde{Q}^{4} E^{2}
\tilde{L} -1)\tilde{r}^{2}+(-\tilde{Q}^{2}
+2)\tilde{r}=R_{El}(\tilde{r}).
\end{align}
Also, in these set of equations, the values of $\tilde{L}$ and $E^{2}$ in the right hand side 
refer to indices of the left hand side of them, that we ignore indices for 
$\tilde{L}$ and $E^{2}$  for simplicity.

\subsection{Analytical solution of geodesic equations}

In this section, we present the solution of the equations of motion
analytically. In the Eqs.(\ref{drt/dfi- GM})-(\ref{drt/dfi- El}) for
the test particle ($\epsilon=1$) and light ray ($\epsilon=0$), we have
polynomials of degree four in the form 
$R(\tilde{r})=\sum_{i=0}^{4} a_{i} \tilde{r}^{i}$, 
with only simple zeros, which for solving them
in this way, we can apply up to two substitutions. The first
substitution is $\tilde{r}=\frac{1}{z} + \tilde{r}_{R}$ , where
$\tilde{r}_{R}$ is a zero of $R$, transforms the problem to
\begin{equation}\label{dz/dfi}
(\dfrac{dz}{d\varphi})^{2}= R_{3}(z) = \sum_{j=1}^{3} b_{j} z^{j}
,\qquad z(\varphi_{0})=z_{0} ,
\end{equation}
with a polynomial $R_{3}$ of degree 3. Where
\begin{equation}
b_{j}=\dfrac{1}{(4-j)!}\dfrac{d^{(4-j)}R}{d \tilde{r}^{(4-j)}} (\tilde{r}_{R}) ,
\end{equation}
in which $b_{j},~(j=1,2,3)$  is an arbitrary constant for each metric which
is related to the parameter of the relevant metric. A
second substitution $z=\frac{1}{b_{3}}(4y-\frac{b_{2}}{3})$, 
changes $R_{3}(z)$, into the Weierstrass form 
\begin{equation}\label{dy/dfi}
(\dfrac{dy}{d\varphi})^{2}=4y^{3}-g_{2}y-g_{3} ,
\end{equation}
where
\begin{equation}
g_{2}=\frac{1}{16}(\frac{4}{3}b_{2}^{2}-4b_{1}b_{3}) , \qquad
g_{3}=\frac{1}{16}(\frac{1}{3}b_{1}b_{2}b_{3}-
\frac{2}{27}b_{2}^{3}-b_{0}b_{3}^{2}) ,
\end{equation}
are the Weierstrass invariants. The differential
equation(\ref{dy/dfi}) is of elliptic type and we used the
Weierstrass $\wp$ function to solve it
\cite{Hackmann:2008zz,Soroushfar:2015wqa}
\begin{equation}
y(\varphi)=\wp (\varphi - \varphi_{in};g_{2},g_{3}),
\end{equation}
where $\varphi_{in}=\varphi_{0}+
\int_{y_{0}}^\infty\dfrac{dy}{\sqrt{4y^{3}-g_{2}y-g_{3}}}$ with
$\varphi_{0}=\frac{1}{4}(\dfrac{b_{3}}{\tilde{r_{0}}-\tilde{r_{R}}}+\dfrac{b_{2}}{3})$
depends only on the initial values $\varphi_{0}$ and
$\tilde{r}_{0}$. Therefore, the solution of Eqs.(\ref{drt/dfi-
GM})-(\ref{drt/dfi- El}) takes the form
\begin{equation} \label{rtfi}
\tilde{r}(\varphi)=\dfrac{b_{3}}{4\wp(\varphi
-\varphi_{in};g_{2},g_{3})-\dfrac{b_{2}}{3}}+\tilde{r}_{R} .
\end{equation}

This is the analytic solution of the equation of motion of a test
particle and light ray in a GMGHS, magnetically charged GMGHS and
electrically charged GMGHS spacetimes. This solution is valid in all
regions of this spacetimes.

\subsection{Orbits}\label{orbits}

In a special spacetime with electric charge, the shape of an orbit 
depends on three paremeters, in which the angular momentum, 
$L$ and the energy, $E$ are the specifications of test particle or 
light ray and the electric charge, $Q$ comes from the related 
spacetime (the mass can be absorbed through a rescaling of 
the radial coordinate). The polynomial $R(r)$ defined
in Eqs.(\ref{dr/dfi- GM})-(\ref{dr/dfi- El}) are included all these
quantities. Since $r$ should be real and positive, the physically
admissible regions are given by those $r$ for which $E^{2} \geq
V_{eff}$ which is presented on the left hand side of
Eqs.(\ref{dr/ds- GM})-(\ref{dr/ds- El}). Therefore, the form of the
resulting orbits are characterized uniquely by the number of
positive real zeros of $R$.

In the following, we introduce different types of orbit. Let $\tilde{r}_+$ be the outer event horizon and $\tilde{r}_-$ be the inner horizon.
\begin{enumerate}	
	\item \textit{Escape orbit} (EO) with range $\tilde{r} \in [r_1, \infty)$ with $r_1>\tilde{r}_+$, or with range $\tilde{r} \in (-\infty, r_1]$ with  $r_1<0$.
	\item \textit{Two-world escape orbit} (TEO) with range $[r_1, \infty)$ where $0<r_1 < r_-$.	
	\item \textit{Bound orbit} (BO) with range $\tilde{r} \in [r_1, r_2]$ with
	\begin{enumerate}
		\item $r_1, r_2  > r_+$, or 
		\item $ 0 < r_1, r_2 < r_-$.
	\end{enumerate}
	\item \textit{Many-world bound orbit} (MBO) with range $\tilde{r} \in [r_1, r_2]$ where $0<r_1 \leq r_-$ and $r_2 \geq r_+$.
	\item \textit{Terminating orbit} (TO) with ranges either $\tilde{r} \in [0, \infty)$ or $\tilde{r} \in [0, r_1]$ with
	\begin{enumerate}
		\item $r_1\geq \tilde{r}_+$, or 
		\item $0<r_1<\tilde{r}_-$.
	\end{enumerate}
\end{enumerate}

Other types of orbits are exceptional and treated separately. They
are connected with the appearance of multiple zeros in $R(r)$ or
with parameter values which reduce the degree of $R(r)$. In both
cases the differential Eqs.(\ref{dr/dfi- GM})-(\ref{dr/dfi- El})
have much simplified structure. These orbits are radial geodesics
with $L = 0$, circular orbits with constant $r$ and orbits
asymptotically approaching circular orbits.

Defining the borders of $R(r) \geq 0$ or, equivalently, $E^{2} \geq
V_{eff}$ is done by the four regular types of geodesic motion
correspond to various arrangements of the real and positive zeros of
$R(r)$. If $R(r)$ has no real and positive zeros, a
terminating orbit is possible if $R(r) > 0$ for all $r > 0$,
but else no geodesic motion is allowed. If $R(r)$ has at least one
real and positive zero then an escape orbit is possible, or a terminating 
orbit if $R(r) > 0$
for $0 < r < r_{1}$, where, $r_{1}$ is the smallest positive zero. If
$R(r)$ has at least two real zeros, $ r_{1} < r_{2}$ with $R(r) > 0$
for $ r_{1} < r < r_{2}$ a bound orbit is permitted. If $R(r)$ is
such that, multiple types of orbits are possible, then the actual orbit
depends on the initial position of the test particle or light ray.

In the following, we will analyse possible types of orbits. 
The major point in this analysis
is that Eqs.(\ref{drt/dfi- GM})-(\ref{drt/dfi- El}) implies
$R(\tilde{r}) \geq 0$, as a necessary condition for the existence of
a geodesic. Thus, the zeros of $R(\tilde{r})$, are extremal values of
$\tilde{r}(\varphi)$ and determine (together with the sign of
$R(\tilde{r})$ between two zeros) the type of geodesic. The
polynomial $R(\tilde{r})$ is in our metrics of degree 4 and,
therefore, has 4 (complex) zeros of which the positive real zeros
are of interest for the type of orbit.

For a given set of parameters $\epsilon,E^{2},\tilde{Q}$, and
$\tilde{L}$ the polynomial $R(\tilde{r})$ has a certain number of
positive real zeros. If $E^{2}$ and $\tilde{L}$ are varied, this
number can change only if two zeros of $R(\tilde{r})$ merge to one.
Solving $R( \tilde{r})=0$ , $\dfrac{dR( \tilde{r})}{d \tilde{r}} =0$
for $E^{2}$ and $\tilde{L}$, for $\epsilon=1$ yields
\begin{align}
E_{GM}^{2}=\dfrac{\tilde{Q}^{2}\tilde{r}^{2}-4\tilde{Q}^{2}\tilde{r}-2\tilde{r}^{3}+4\tilde{Q}^{2}+8\tilde{r}^{2}-8\tilde{r}}{(\tilde{Q}^{2}\tilde{r}-4\tilde{Q}^{2}-2\tilde{r}^{2}+6\tilde{r})\tilde{r}}
,\qquad
\tilde{L}_{GM}=-\frac{1}{2}\dfrac{\tilde{Q}^{2}\tilde{r}-4\tilde{Q}^{2}-2\tilde{r}^{2}+6\tilde{r}}{\tilde{r}(\tilde{Q}^{4}-2\tilde{Q}^{2}\tilde{r}+\tilde{r}^{2})},
\end{align}
\begin{align}
E_{Mag}^{2}=-\dfrac{2(\tilde{r}^{2}-4\tilde{r}+4)}{\tilde{Q}^{2}\tilde{r}-4\tilde{Q}^{2}-2\tilde{r}^{2}+6\tilde{r}} ,\qquad
\tilde{L}_{Mag}=\dfrac{\tilde{Q}^{2}\tilde{r}-4\tilde{Q}^{2}-2\tilde{r}^{2}+6\tilde{r}}{\tilde{r}^{3}(\tilde{Q}^{2}-2)},
\end{align}
\begin{align}
E_{El}^{2}=\dfrac{2(\tilde{Q}^{4}+2\tilde{Q}^{2}\tilde{r}-4\tilde{Q}^{2}+\tilde{r}^{2}-4\tilde{r}+4)\tilde{r}}{(\tilde{Q}^{2}+\tilde{r})(\tilde{Q}^{4}+3\tilde{Q}^{2}\tilde{r}-2\tilde{Q}^{2}+2\tilde{r}^{2}-6\tilde{r})}
,\qquad
\tilde{L}_{El}=\dfrac{\tilde{Q}^{4}+3\tilde{Q}^{2}\tilde{r}-2\tilde{Q}^{2}+2\tilde{r}^{2}-6\tilde{r}}{\tilde{r}^{2}(\tilde{Q}^{4}+\tilde{Q}^{2}\tilde{r}-2\tilde{Q}^{2}+2\tilde{r})}.
\end{align}

\begin{figure}[h]
	\centering	
	\subfigure[]{
		\includegraphics[width=0.4\textwidth]{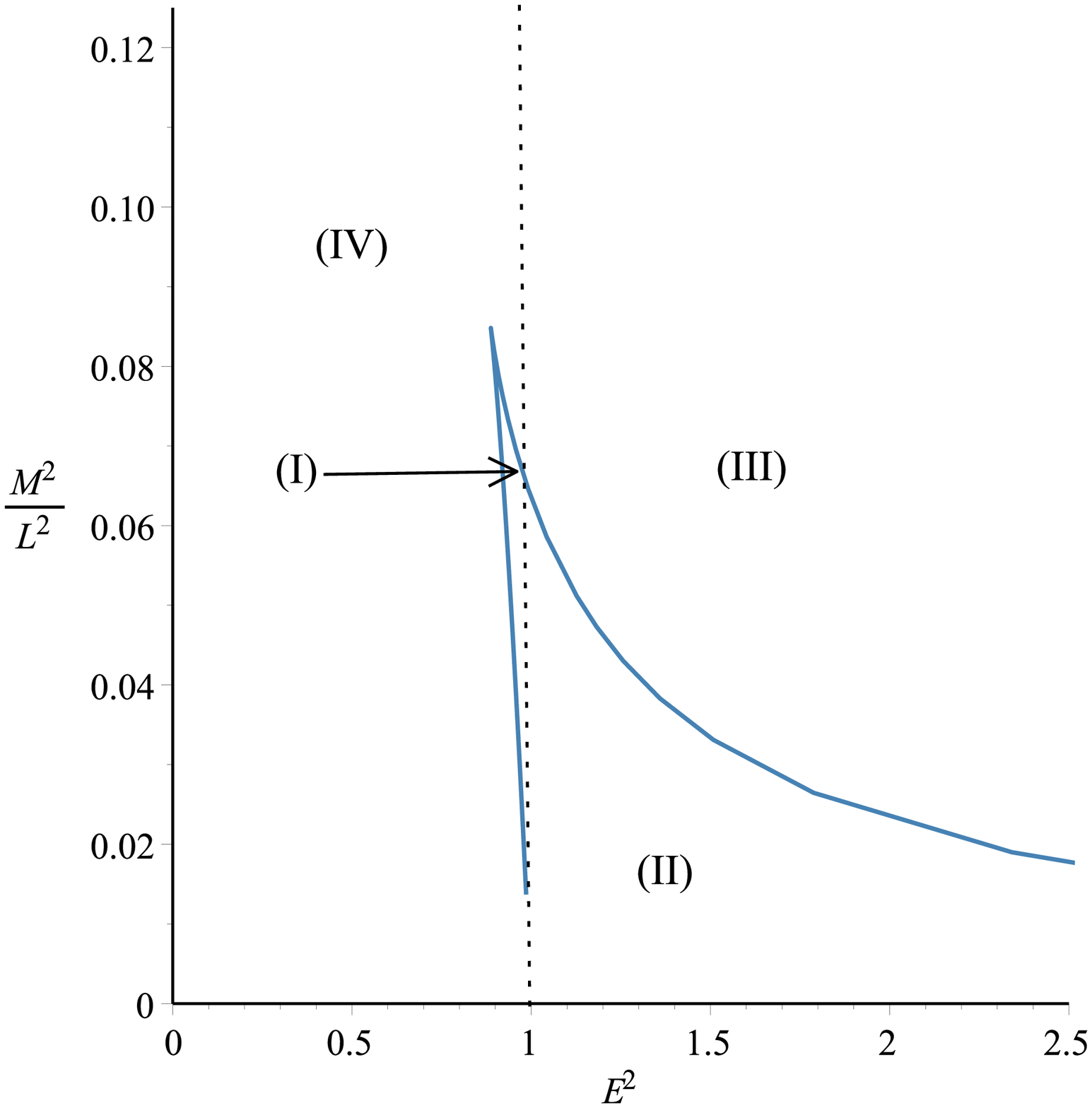}
	}
	\subfigure[]{
		\includegraphics[width=0.4\textwidth]{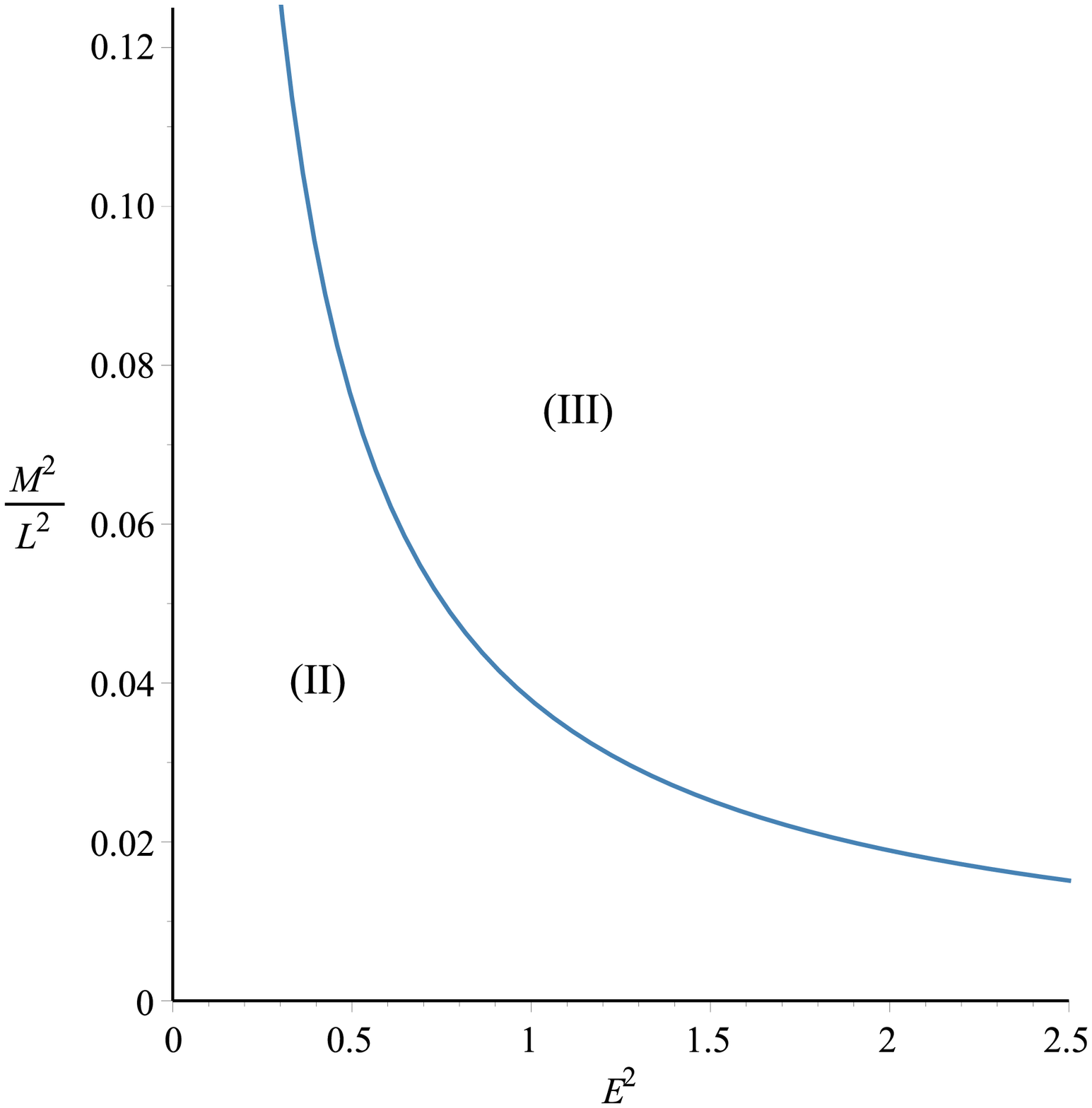}
	}
	\caption{ \label{LE2} Regions of different types of geodesic
motion for test particles, in GMGHS, magnetically
charged and electrically charged GMGHS spacetimes. 
$\tilde{Q}=0.25 $, $\epsilon = 1$ and $\epsilon = 0$, for (a) and (b) respectively. 
The numbers of positive real zeros in these regions are: I=4, II=3, III=1 and IV=2, for GMGHS, magnetically
charged GMGHS spacetimes and I=3, II=2, III=0 and IV=1, for electrically charged GMGHS spacetime.}
\end{figure}

In Fig.\ref{LE2}(a), the result of this analysis is shown for test
particles ($\epsilon =1$). For light rays $(\epsilon = 0)$, 
the analysis is the same as in the $(\epsilon = 1)$ case and the result of this analysis is 
shown in Fig.\ref{LE2}(b). 
 
%\clearpage

Four different regions can be
identified in Fig.\ref{LE2}, for timelike geodesic but for null
geodesic, in Fig.\ref{LE2}(b), we have only two regions.
It should be noted that, for electrically charged GMGHS metric, the
number of zeros is one more less than zeros of GMGHS and magnetically
charged GMGHS metrics in each region. Summary of possible orbit types can be
found in Tabels~\ref{tab:GMG} and~\ref{tab:EL}.
%%%%%%%%%%%%%%%%%%%%
\begin{table}[h]
\begin{center}
\begin{tabular}{|l|l|c|l|}
%{|lccll|}
\hline
region & pos.zeros  & range of $\tilde{r}$ & orbit \\
\hline\hline
I & 4 &
\begin{pspicture}(-3,-0.2)(2.2,0.2)%\psgrid
\psline[linewidth=0.5pt]{->}(-2.5,0)(1.75,0)
\psline[linewidth=0.5pt](-2.5,-0.2)(-2.5,0.2)
\psline[linewidth=0.5pt,doubleline=true](-2.25,-0.2)(-2.25,0.2)
\psline[linewidth=0.5pt,doubleline=true](-1.25,-0.2)(-1.25,0.2)
%\psline[linewidth=1.2pt]{-*}(-4,0)(-3,0)
\psline[linewidth=1.2pt]{*-*}(-2.25,0)(-0.75,0)
\psline[linewidth=1.2pt]{*-*}(0.25,0)(1.25,0)
\end{pspicture}
  & MBO, BO
\\  \hline
II & 3 &
\begin{pspicture}(-3,-0.2)(2.2,0.2)%\psgrid
\psline[linewidth=0.5pt]{->}(-2.5,0)(1.75,0)
\psline[linewidth=0.5pt](-2.5,-0.2)(-2.5,0.2)
\psline[linewidth=0.5pt,doubleline=true](-2.25,-0.2)(-2.25,0.2)
\psline[linewidth=0.5pt,doubleline=true](-1.25,-0.2)(-1.25,0.2)
\psline[linewidth=1.2pt]{*-*}(-2.25,0)(-0.75,0)
\psline[linewidth=1.2pt]{*-}(0.25,0)(1.75,0)
\end{pspicture}
  & EO و MBO
\\ \hline
III  & 1 &
\begin{pspicture}(-3,-0.2)(2.2,0.2)%\psgrid
\psline[linewidth=0.5pt]{->}(-2.5,0)(1.75,0)
\psline[linewidth=0.5pt](-2.5,-0.2)(-2.5,0.2)
\psline[linewidth=0.5pt,doubleline=true](-2.25,-0.2)(-2.25,0.2)
\psline[linewidth=0.5pt,doubleline=true](-1.25,-0.2)(-1.25,0.2)
%\psline[linewidth=1.2pt]{-*}(-4,0)(-3,0)
\psline[linewidth=1.2pt]{*-}(-2.25,0)(1.75,0)
\end{pspicture}
& TEO
\\ \hline
IV & 2 &
\begin{pspicture}(-3,-0.2)(2.2,0.2)%\psgrid
\psline[linewidth=0.5pt]{->}(-2.5,0)(1.75,0)
\psline[linewidth=0.5pt](-2.5,-0.2)(-2.5,0.2)
\psline[linewidth=0.5pt,doubleline=true](-2.25,-0.2)(-2.25,0.2)
\psline[linewidth=0.5pt,doubleline=true](-1.25,-0.2)(-1.25,0.2)
%\psline[linewidth=1.2pt]{-*}(-4,0)(-3,0)
%\psline[linewidth=1.2pt]{*-*}(-1,0)(1.5,0)
\psline[linewidth=1.2pt]{*-*}(-2.25,0)(-0.75,0)
\end{pspicture}
  & MBO

\\ \hline\hline
\end{tabular}
\caption{Types of orbits of GMGHS and magnetically GMGHS black
holes. The range of the orbits is represented by thick lines. The
dots show the turning points of the orbits.
The positions of the two horizons are marked by a vertical double line.
 The single vertical line indicates $\tilde{r}=0$.}
\label{tab:GMG}
\end{center}
\end{table}
%%%%%%%%%%%%%%%%%%%%%
\begin{table}[h]
\begin{center}
\begin{tabular}{|l|l|c|l|}
%{|lccll|}
\hline
region & pos.zeros  & range of $\tilde{r}$ & orbit \\
\hline\hline
I & 3 &
\begin{pspicture}(-3,-0.2)(2.2,0.2)%\psgrid
\psline[linewidth=0.5pt]{->}(-2.5,0)(1.75,0)
\psline[linewidth=0.5pt](-2.5,-0.2)(-2.5,0.2)
%\psline[linewidth=0.5pt,doubleline=true](-2.25,-0.2)(-2.25,0.2)
\psline[linewidth=0.5pt,doubleline=true](-1.25,-0.2)(-1.25,0.2)
%\psline[linewidth=1.2pt]{-*}(-4,0)(-3,0)
\psline[linewidth=1.2pt]{-*}(-2.5,0)(-0.75,0)
\psline[linewidth=1.2pt]{*-*}(0.25,0)(1.25,0)
\end{pspicture}
  & TO, BO
\\  \hline
II & 2 &
\begin{pspicture}(-3,-0.2)(2.2,0.2)%\psgrid
\psline[linewidth=0.5pt]{->}(-2.5,0)(1.75,0)
\psline[linewidth=0.5pt](-2.5,-0.2)(-2.5,0.2)
%\psline[linewidth=0.5pt,doubleline=true](-2.25,-0.2)(-2.25,0.2)
\psline[linewidth=0.5pt,doubleline=true](-1.25,-0.2)(-1.25,0.2)
\psline[linewidth=1.2pt]{-*}(-2.5,0)(-0.75,0)
\psline[linewidth=1.2pt]{*-}(0.25,0)(1.75,0)
\end{pspicture}
  & TO, EO
\\ \hline
III  & 0 &
\begin{pspicture}(-3,-0.2)(2.2,0.2)%\psgrid
\psline[linewidth=0.5pt]{->}(-2.5,0)(1.75,0)
\psline[linewidth=0.5pt](-2.5,-0.2)(-2.5,0.2)
%\psline[linewidth=0.5pt,doubleline=true](-2.25,-0.2)(-2.25,0.2)
\psline[linewidth=0.5pt,doubleline=true](-1.25,-0.2)(-1.25,0.2)
%\psline[linewidth=1.2pt]{-*}(-4,0)(-3,0)
%\psline[linewidth=1.2pt]{*-*}(-1.8,0)(-1,0)
\psline[linewidth=1.2pt]{-}(-2.5,0)(1.75,0)
\end{pspicture}
& TO
\\ \hline
IV & 1 &
\begin{pspicture}(-3,-0.2)(2.2,0.2)%\psgrid
\psline[linewidth=0.5pt]{->}(-2.5,0)(1.75,0)
\psline[linewidth=0.5pt](-2.5,-0.2)(-2.5,0.2)
%\psline[linewidth=0.5pt,doubleline=true](-2.25,-0.2)(-2.25,0.2)
\psline[linewidth=0.5pt,doubleline=true](-1.25,-0.2)(-1.25,0.2)
%\psline[linewidth=1.2pt]{-*}(-4,0)(-3,0)
%\psline[linewidth=1.2pt]{*-*}(-1,0)(1.5,0)
\psline[linewidth=1.2pt]{-*}(-2.5,0)(-0.75,0)
\end{pspicture}
  & TO

\\ \hline\hline
\end{tabular}
\caption{Types of orbits of electrically GMGHS black hole. The range
of the orbits is represented by thick lines. The dots show the
turning points of the orbits.
 The positions of the one horizon is marked by a vertical double line.
 The single vertical line indicates $\tilde{r}=0$.}
\label{tab:EL}
\end{center}
\end{table}
%%%%%%%%%%%%%%%%%%%

\begin{figure}[h]
	\centering
	\subfigure[]{
		\includegraphics[width=0.45\textwidth]{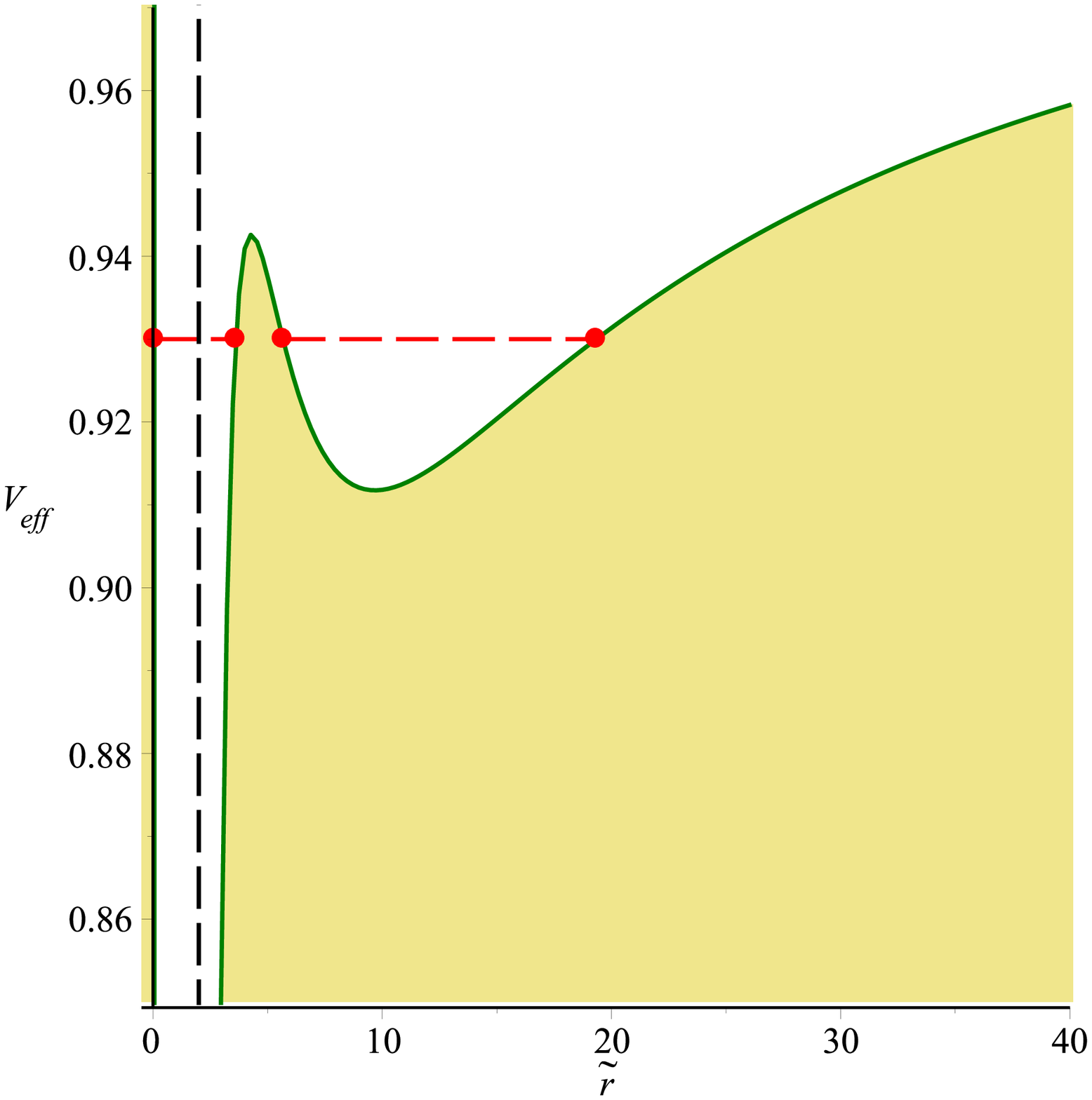}
	}
	%\subfigure[closeup of figure (a)]{
		%\includegraphics[width=0.33\textwidth]{V/VmIb.eps}
	%}
	\subfigure[]{
		\includegraphics[width=0.45\textwidth]{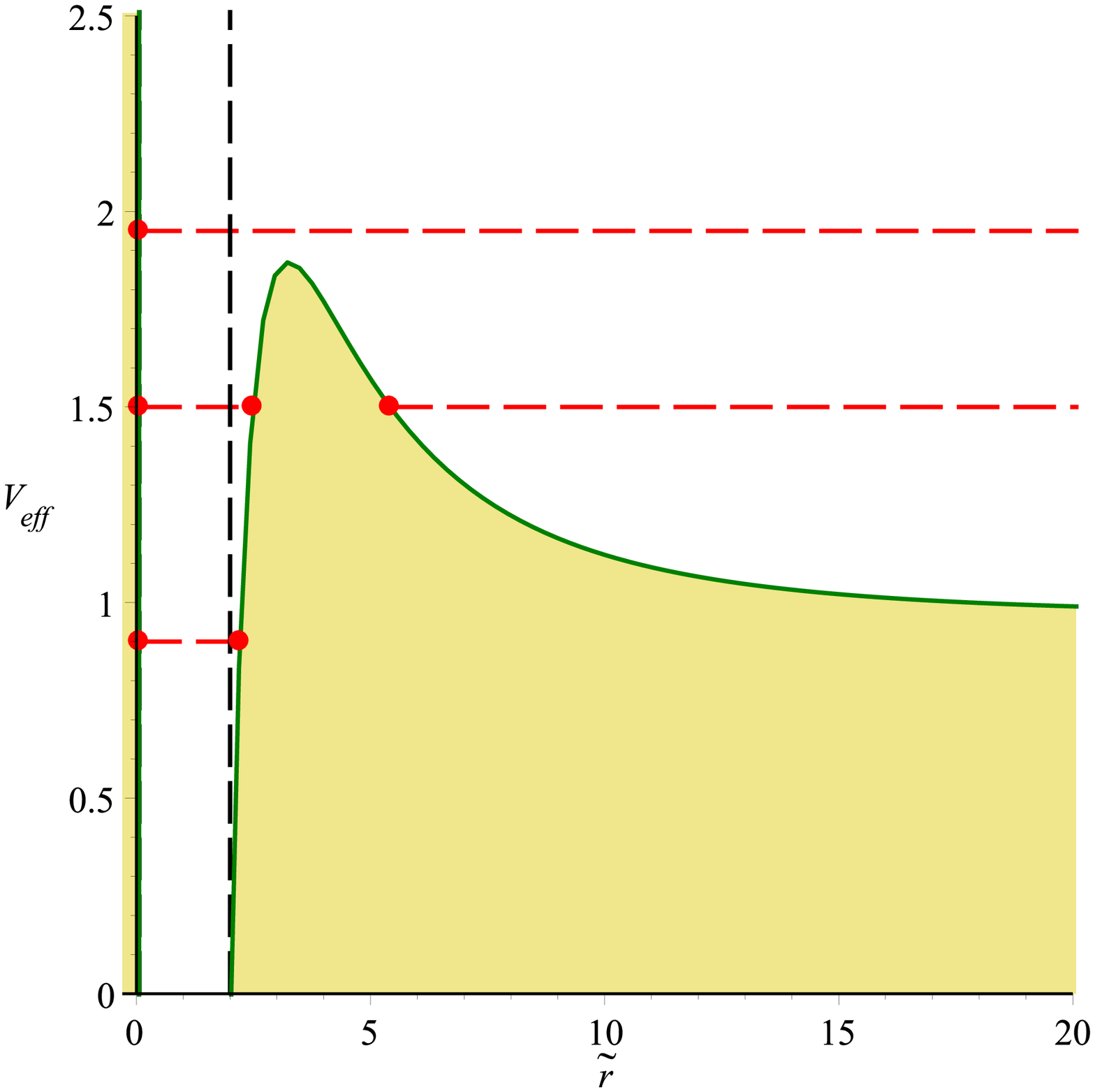}
	}
	%\subfigure[closeup of figure (c)]{%
		%\includegraphics[width=0.33\textwidth]{V/Vmtotb.eps}
	%}
	\subfigure[]{
		\includegraphics[width=0.45\textwidth]{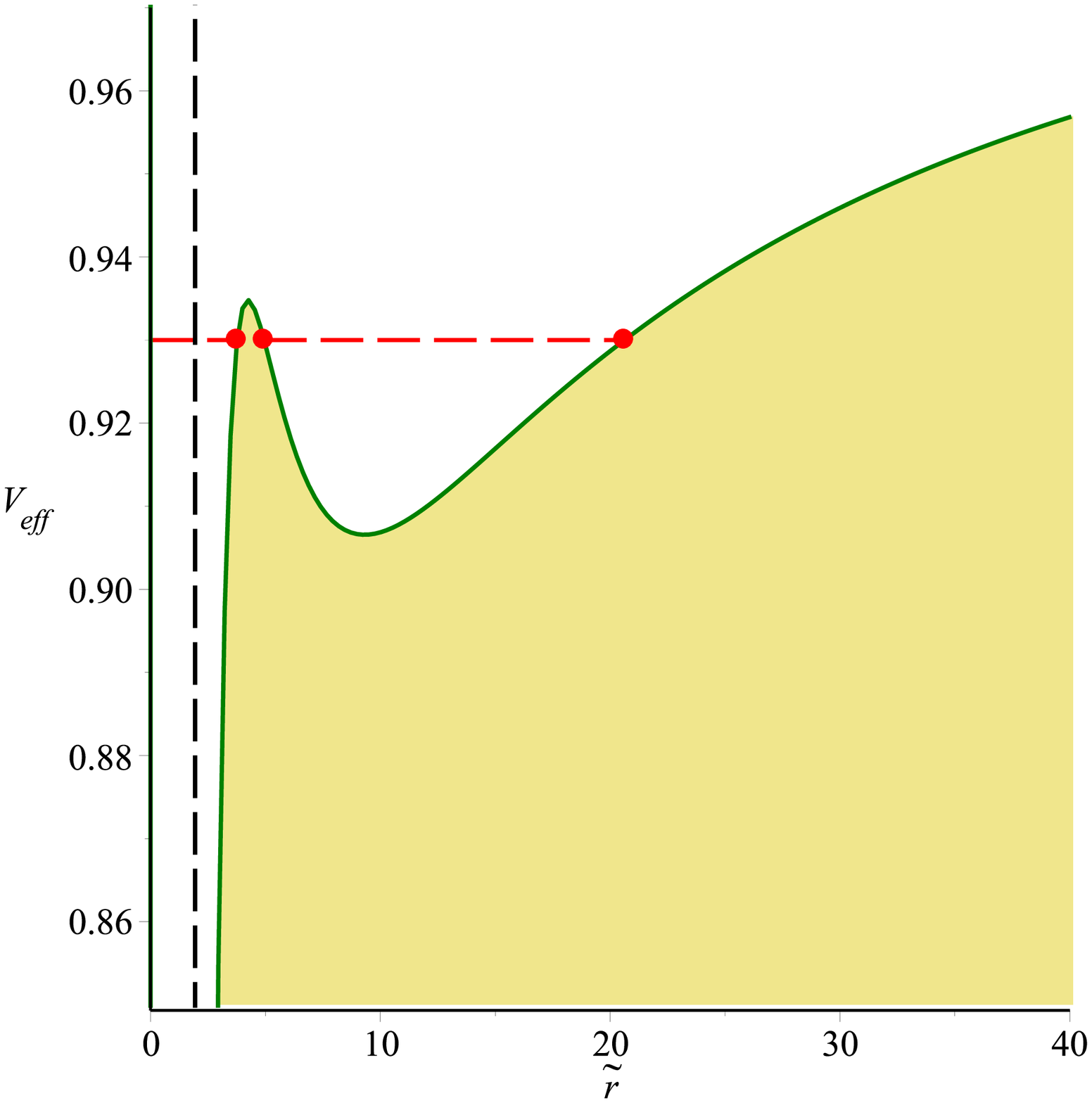}
	}
	\subfigure[]{
		\includegraphics[width=0.45\textwidth]{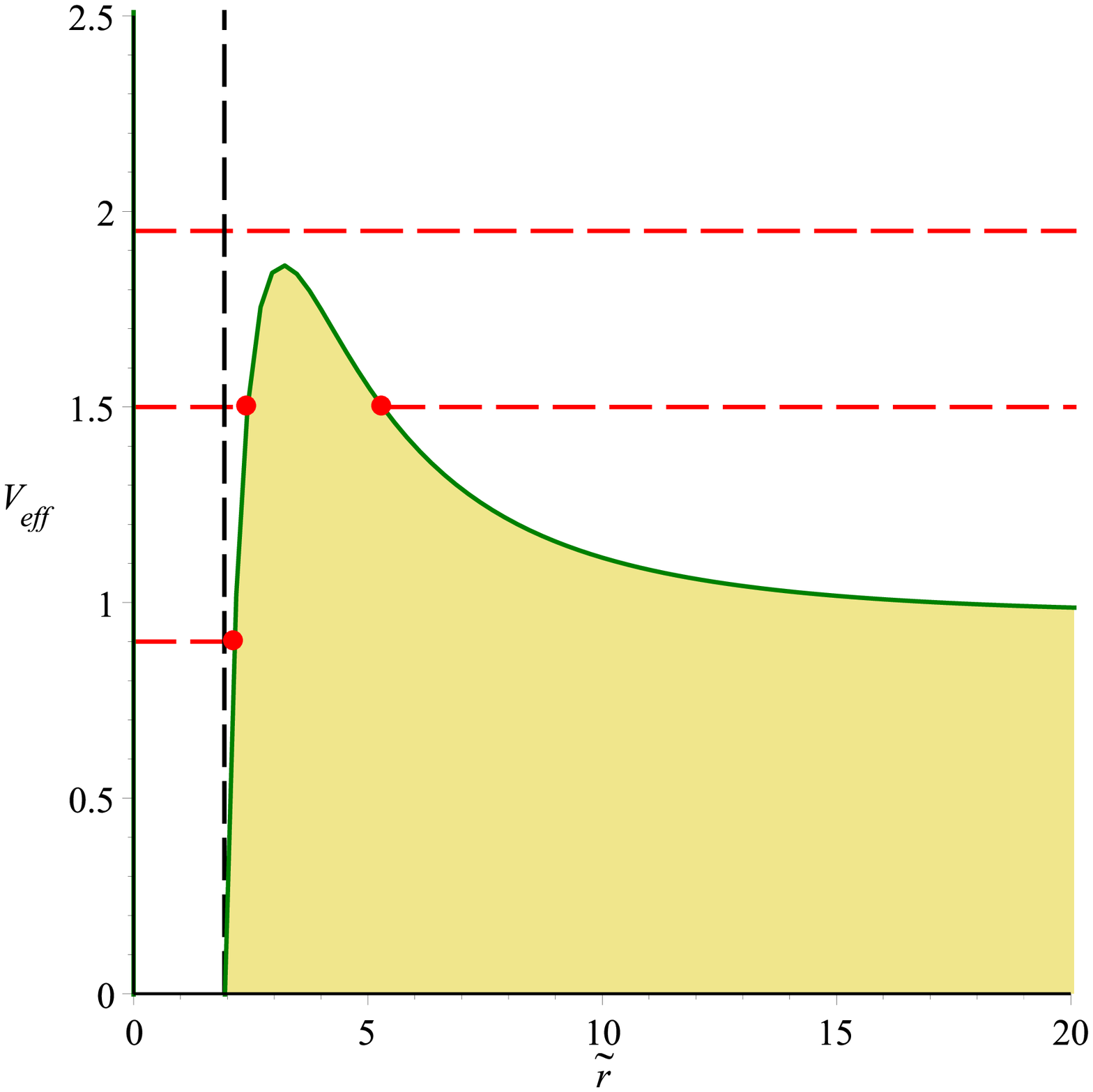}
	}
	\caption{ \label{pic:potential.EMGM} Plots of the effective potential together with examples of energies for the different orbit types. (a), (b) correspond to Table~\ref{tab:GMG} and (c), (d) correspond to Table~\ref{tab:EL}. The green curves represent the effective potential. The red dashed lines correspond to energies. The red dots mark the zeros of the polynomial $R$, which are the turning points of the orbits. In the khaki area no motion is possible since $\tilde{R} <0$. The vertical black dashed lines show the position of the horizons. $\epsilon=1$, $\tilde{Q}=0.25$, and $\tilde{L}=0.072$ for (a), (c) and $\tilde{L}=0.025$ for (b), (d).}
\end{figure}

\begin{figure}
 \centering
 \subfigure[]{
  \includegraphics[width=0.45\textwidth]{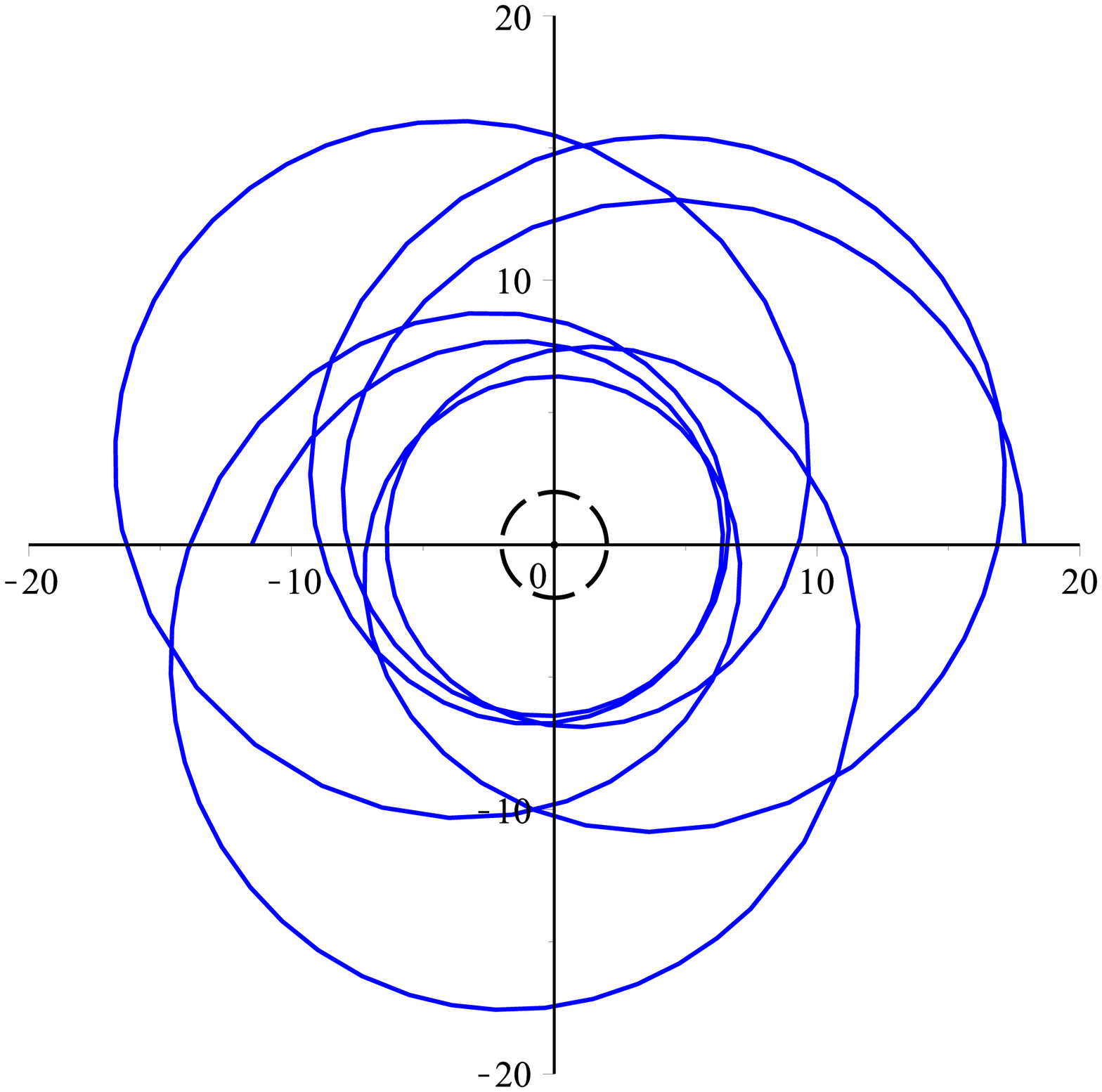}
 }
 \subfigure[]{
  \includegraphics[width=0.45\textwidth]{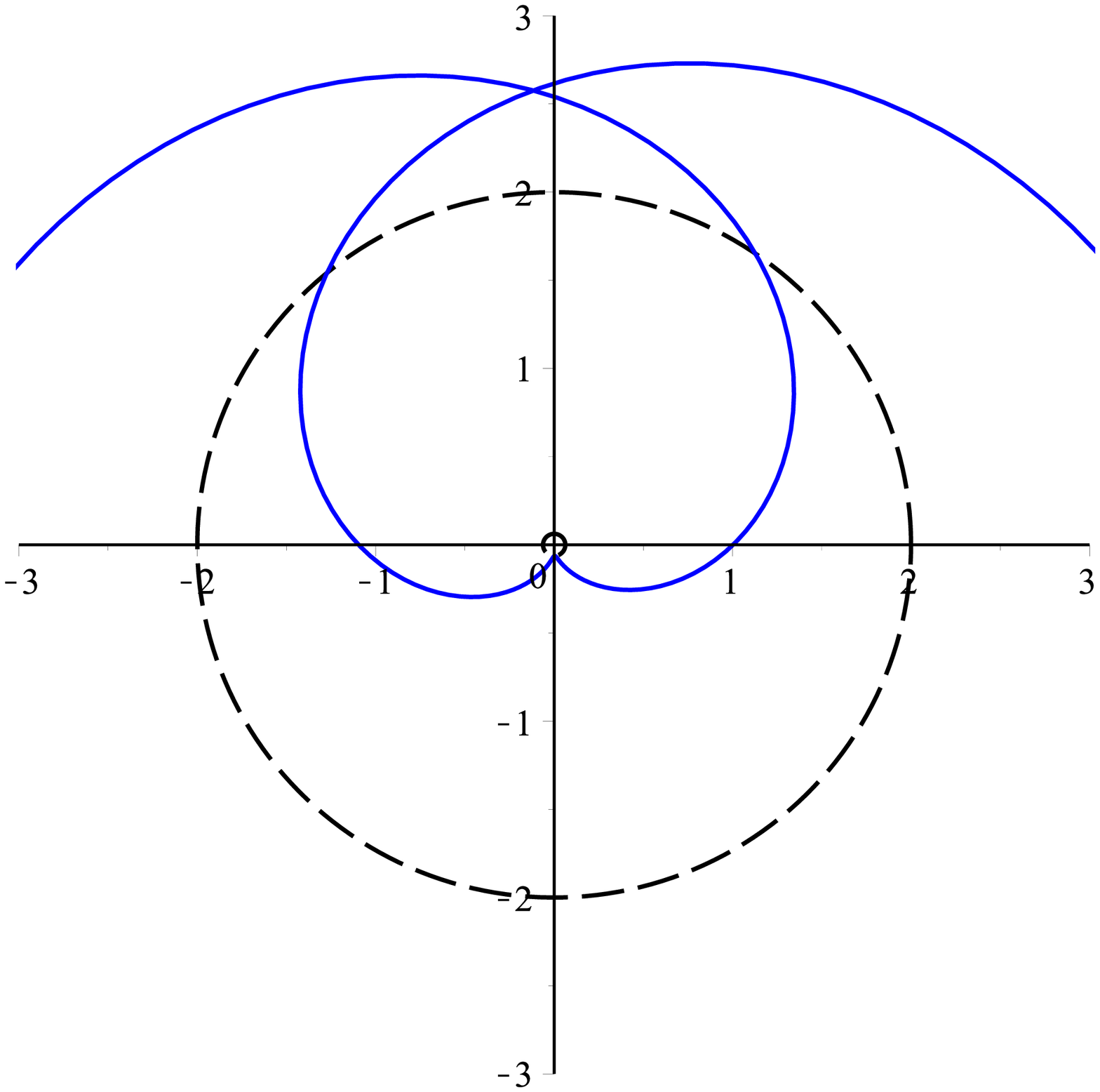}
 }
\subfigure[]{
  \includegraphics[width=0.45\textwidth]{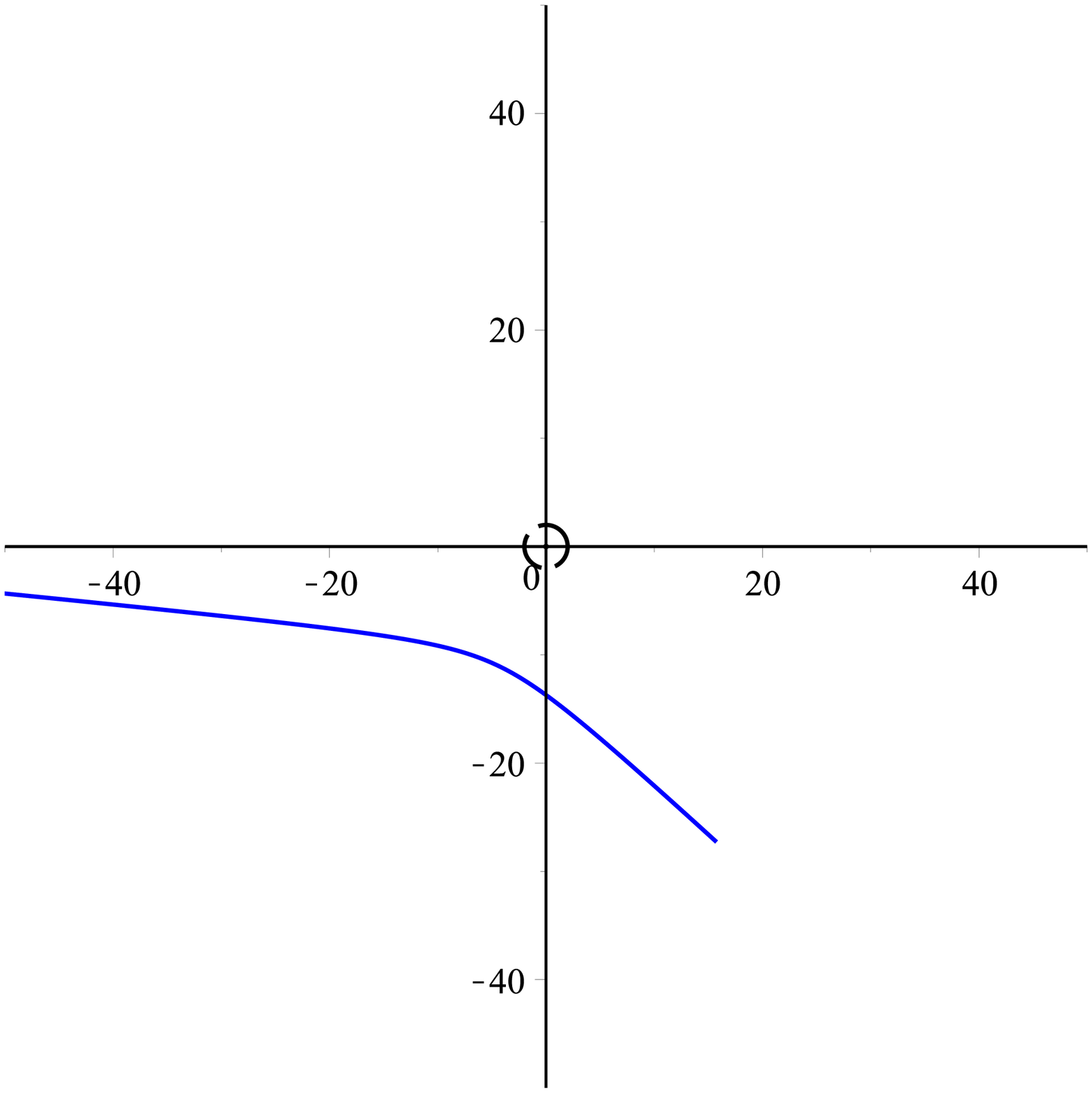}
 }
\subfigure[]{
  \includegraphics[width=0.45\textwidth]{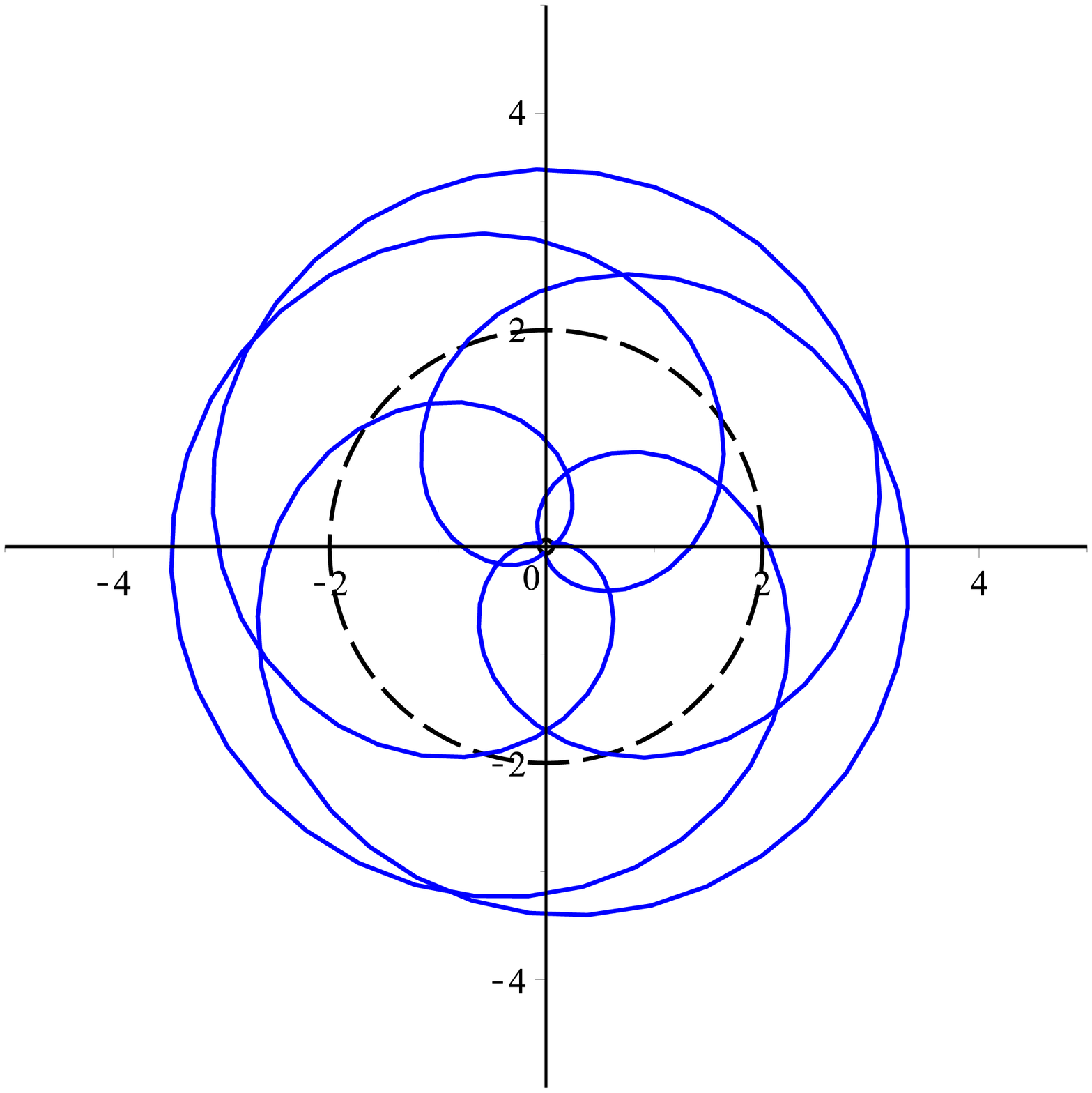}
 }
 \caption{Four examples of possible orbits in the spacetime of GMGHS and magnetically charged GMGHS black holes. A bound orbit (a), a two-world escape orbit (c), an escape orbit and a Mani-world bound orbit (d), with parameters $\epsilon=1$, $\tilde{Q}=0.25$ and $L=0.072$, $E=\sqrt{0.93}$ for (a), (d) and $L=0.025$, $E=\sqrt{1.94}$ for (b) and $L=0.025$, $E=\sqrt{1.5}$ for (c). The blue lines show the path of the orbits and the circles represent the inner and outer horizons.}
\label{pic:Orbits.MGM}
\end{figure}

\begin{figure}
 \centering
  \subfigure[]{
  \includegraphics[width=0.47\textwidth]{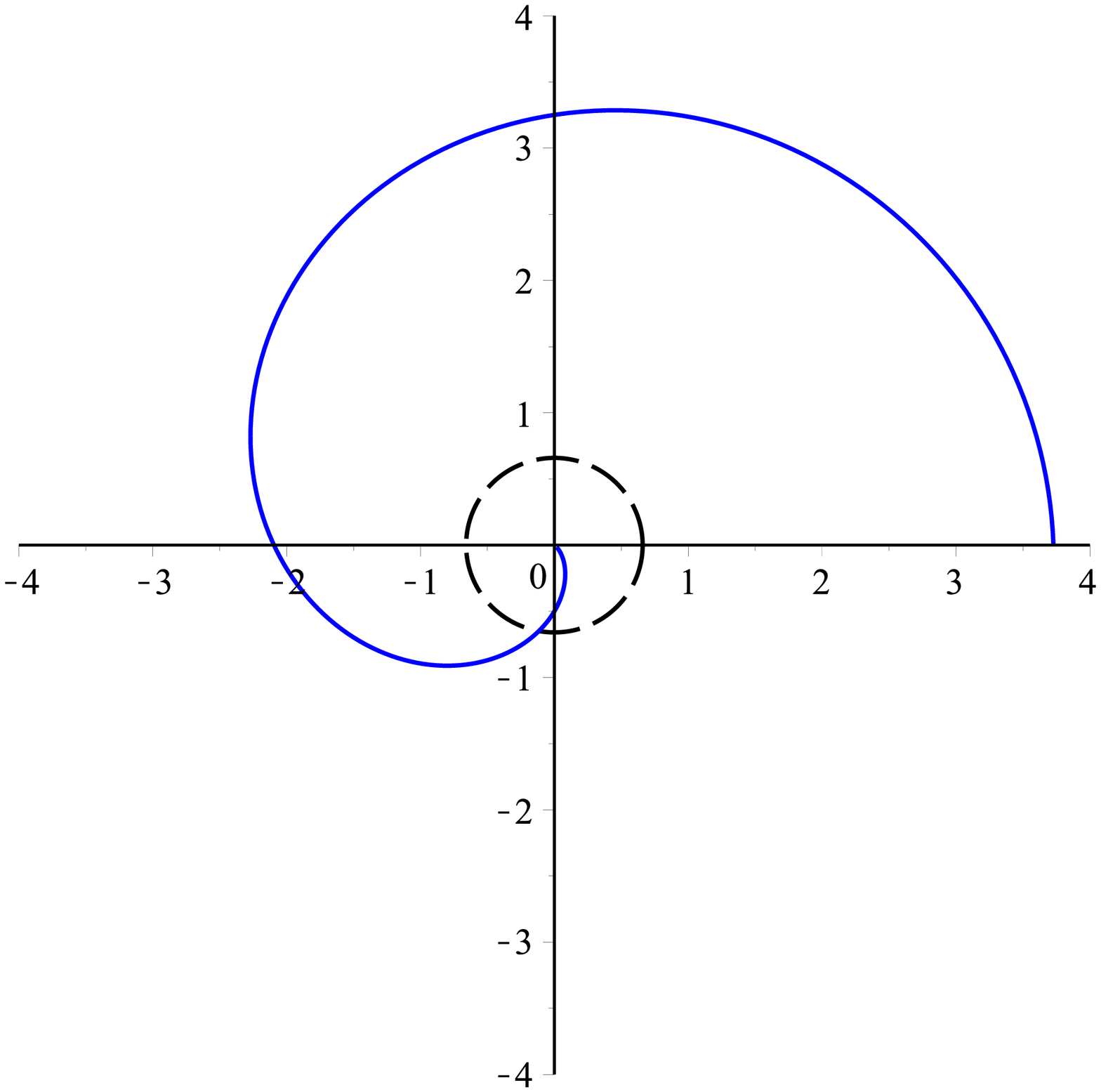}
 }
 \subfigure[]{
  \includegraphics[width=0.47\textwidth]{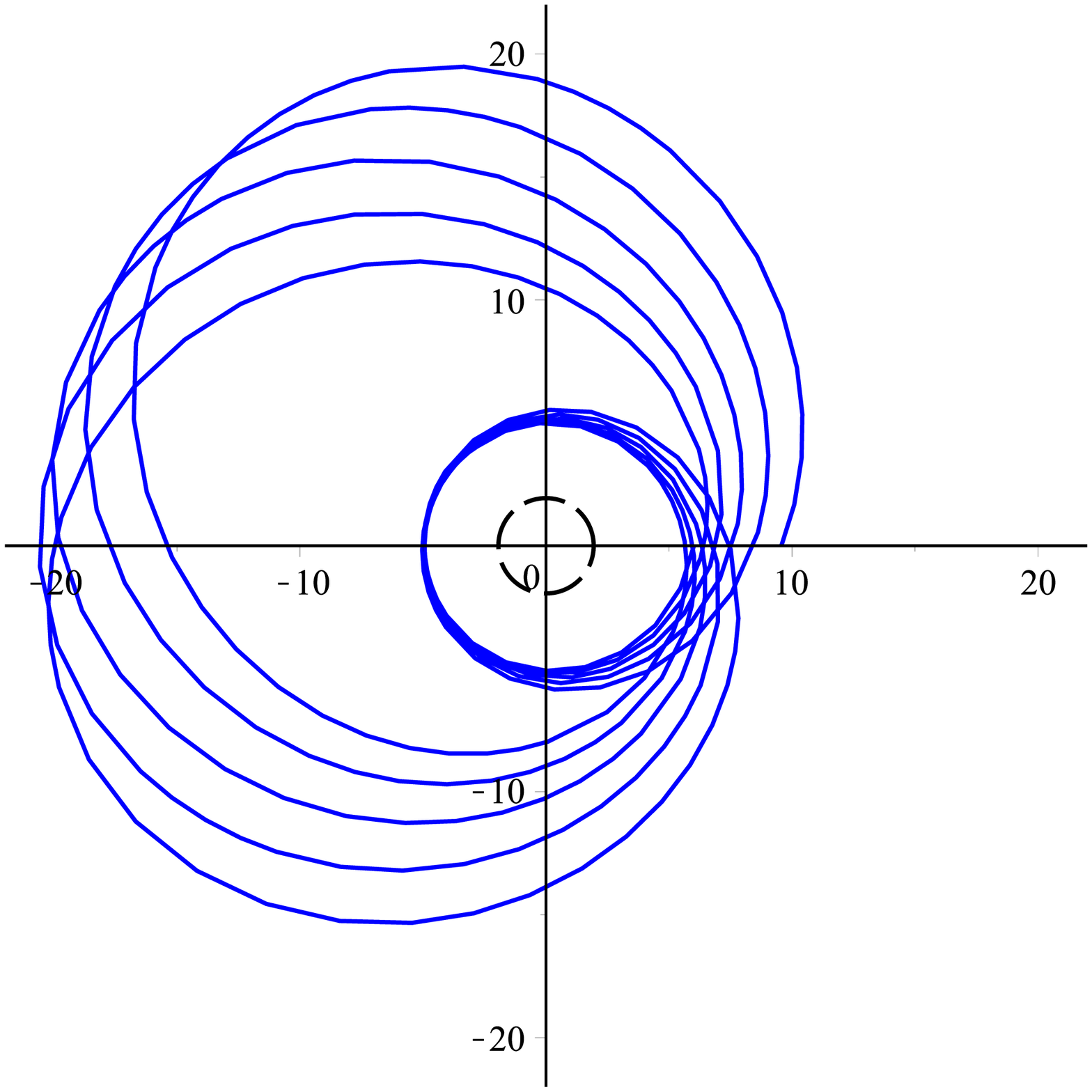}
 }
\subfigure[]{
  \includegraphics[width=0.47\textwidth]{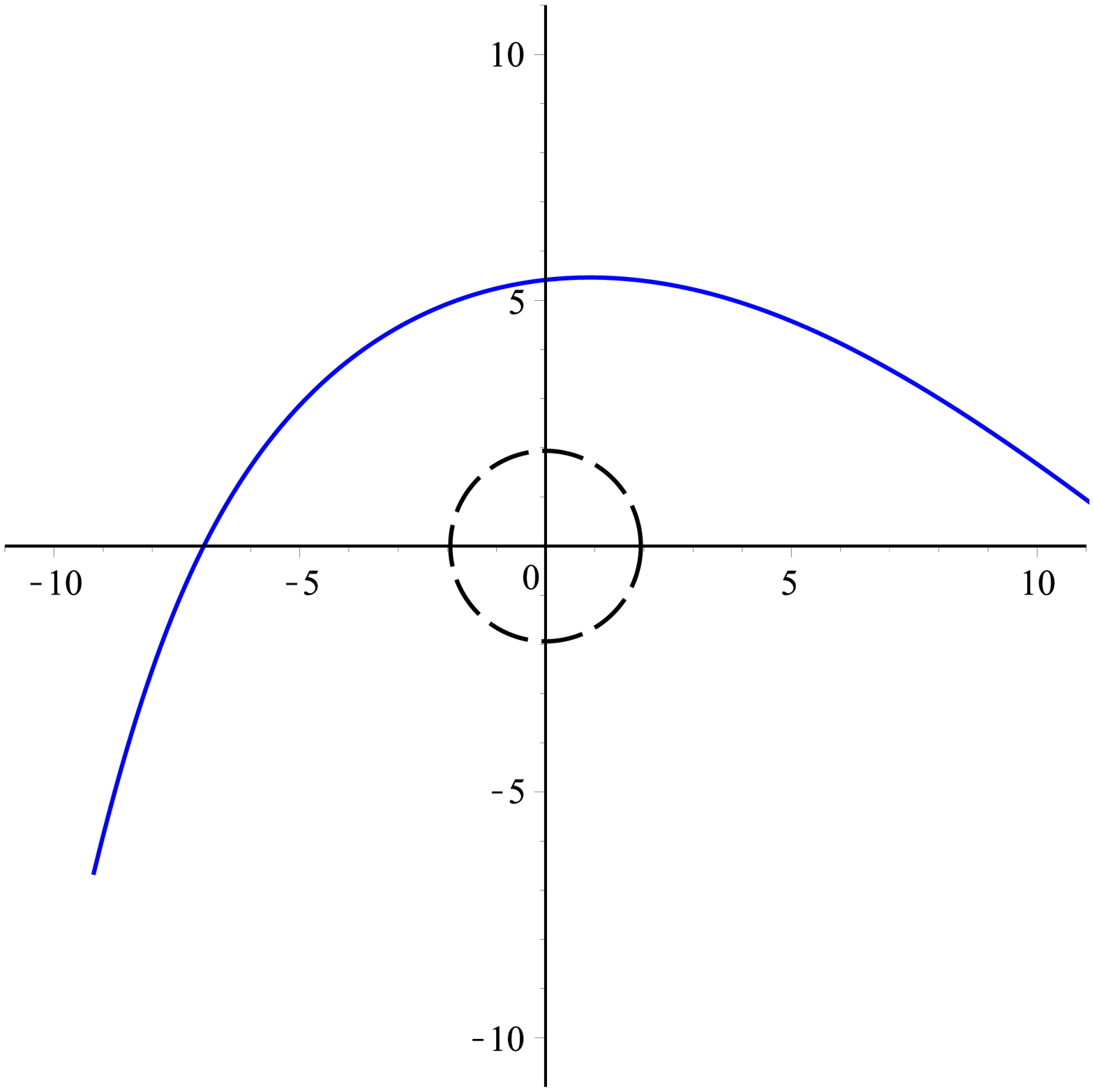}
 }
 
 \caption{Three examples of possible orbits in the spacetime of a electrically charged GMGHS black hole. A bound orbit (a), a terminating orbit (b) and an escape orbit (c), with parameters $\epsilon=1$, $\tilde{Q}=0.25$ and $L=0.072$, $E=\sqrt{0.93}$ for (a), $L=0.025$, $E=\sqrt{1.94}$ for (b) and $L=0.025$, $E=\sqrt{1.5}$ for (c). The blue lines show the path of the orbits and the circle represents the event horizon.}
\label{pic:Orbits.EL}
\end{figure}

In the following, we give a list of existence regions (for Fig.~\ref{LE2} and tables~\ref{tab:GMG} and \ref{tab:EL}).
For each region, examples of effective potentials and possible orbit types are demonstrated in Figs.~\ref{pic:potential.EMGM}--\ref{pic:Orbits.EL}.

\begin{enumerate}
\item 
In region I, for GMGHS and magnetically charged GMGHS black holes, $R(\tilde{r})$ has 4 positive
real zeros $r_{1} < r_{2} < r_{3} < r_{4}$ 
with $R(\tilde{r}) \geq 0$ for $r_{1} \leq \tilde{r} \leq r_{2}$ 
and $r_{3} \leq \tilde{r} \leq r_{4}$.
Therefore the possible orbit types are bound and Mani-world bound orbits, respectively [see 
Figs.~\ref{pic:potential.EMGM} (a), \ref{pic:Orbits.MGM} (a), and \ref{pic:Orbits.MGM} (d)]. 
But in this region for electrically charged GMGHS black hole,
$R(\tilde{r})$ has 3 positive real zeros $r_{1} < r_{2} < r_{3} $
with $R(\tilde{r}) \geq 0$ for $0 \leq \tilde{r} \leq r_{1}$ and
$r_{2} \leq \tilde{r} \leq r_{3} $. Therefore the possible orbit types are terminating and
 bound orbits, respectively [see Figs.~\ref{pic:potential.EMGM} (c), \ref{pic:Orbits.EL} (a), and \ref{pic:Orbits.EL} (b)].

\item 
In region II, for GMGHS and magnetically charged GMGHS black holes, $R(\tilde{r})$ has 3 positive
real zeros $r_{1} < r_{2} < r_{3} $ with $R(\tilde{r}) \geq 0$ for
$r_{1} \leq \tilde{r} \leq r_{2}$ and $r_{3} \leq \tilde{r}$.
Therefore the possible orbit types are Mani-world bound and escape orbits, respectively [see
Figs.~\ref{pic:potential.EMGM} (b), \ref{pic:Orbits.MGM} (d), and \ref{pic:Orbits.MGM} (c)].
But in this region for electrically charged GMGHS black hole,
$R(\tilde{r})$ has 2 positive real zeros $r_{1}<r_{2}$ with
$R(\tilde{r})>0$ for $0\leq \tilde{r} \leq r_{1}$ and $r_{2} \leq \tilde{r}$. 
Therefore the possible orbit types are terminating and escape orbits, respectively
[see Figs.~\ref{pic:potential.EMGM} (d), \ref{pic:Orbits.EL} (a) and \ref{pic:Orbits.EL} (c)].

\item 
In region III, for GMGHS and magnetically charged GMGHS black holes, $R(\tilde{r})$ has 1
positive real zero $r_{1}$ with $R(\tilde{r})\geq0$ for $r_{1}\leq \tilde{r}$. 
Therefore the possible orbit type is the two-world escape orbit [see
Figs.~\ref{pic:potential.EMGM} (b) and \ref{pic:Orbits.MGM} (b)]. 
But in this region for electrically charged GMGHS black hole,
$R(\tilde{r})$ has 0 positive real zeros and $R(\tilde{r})\geq 0$
for $0\leq \tilde{r}$. Therefore the possible orbit type is terminating 
orbit [see Figs.~\ref{pic:potential.EMGM} (d) and \ref{pic:Orbits.EL} (a)].

\item 
In region IV, for GMGHS and magnetically charged GMGHS black holes, $R(\tilde{r})$ has 2 positive
real zeros $r_{1}$,$r_{2}$ with $R(\tilde{r})\geq0$ for $r_{1}\leq \tilde{r}\leq r_{2}$. 
Therefore the possible orbit type is Mani-world bound orbit [see
Figs.~\ref{pic:potential.EMGM} (b) and \ref{pic:Orbits.MGM} (d)].  
But in this region for electrically charged GMGHS black hole, for electrically charged GMGHS black hole,
$R(\tilde{r})$ has 1 positive real zero $r_{1}$ with
$R(\tilde{r})\geq0$ for positive $r$. Therefore the possible orbit type is
terminating orbit [see Figs.~\ref{pic:potential.EMGM} (d) and \ref{pic:Orbits.EL} (a)].

\end{enumerate}

%%%%%%%%%%%
\clearpage

\subsection{Astrophysical applications}

\subsubsection{\textbf{Deflection of light}}
The deflection of light in a Schwarzschild de Sitter spacetime was discussed by Rindler and Ishak \cite{Rindler:2007zz}. 
Though the equation of motion is the same as in Schwarzschild
spacetime for identical periapsis $ r_{p} $, they showed that the measuring process for angles
reintroduces the effect of the cosmological constant. Also, this method was applied in a more general solution 
of Weyl conformal gravity in Refs.~\cite{Bhattacharya:2009rv},~\cite{Bhattacharya:2010xh}.

According to their scheme, they used the invariant 
formula for the cosine of the angle between two coordinate directions $ d $ and $ \delta $, as
\begin{align} \label{1}
\cos(\psi) = \dfrac{g_{ij} d^{i} \delta^{j}}{(g_{ij} d^{i} d^{j})^{\frac{1}{2}}(g_{ij} \delta^{i} \delta^{j})^{\frac{1}{2}}}.
\end{align}
For our purpose the relevant $ g_{ij} $ is
\begin{align}
(g_{rr})_{GM}= (1-\frac{2M}{r})^{-1} ,      \qquad (g_{\varphi \varphi})_{GM} = r(r-\frac{Q^{2}}{M}),\nonumber\\
(g_{rr})_{Mag}=\Big( (1-\dfrac{2M}{r})(1-\dfrac{Q^{2}}{Mr}) \Big)^{-1},      \qquad (g_{\varphi \varphi})_{Mag} = r^{2},\nonumber\\
(g_{rr})_{El}=\Big( 1+\dfrac{Q^{2}-2M^2}{Mr} \Big)^{-1},      \qquad (g_{\varphi \varphi})_{Mag} = r^{2}.
\end{align}
Then, if we call the direction of the orbit $ d $ and that of the coordinate line $ \varphi = const  \delta $ , we have
\begin{eqnarray}
& d & = (dr , d\varphi) = (\dfrac{dr}{d\varphi}, 1) d\varphi , \qquad (d\varphi < 0) ,\\
& \delta & = (\delta r, 0) = (1, 0)\delta r.
\end{eqnarray}
Substituted into (\ref{1}), these values yield
\begin{align}
\cos(\psi) = \dfrac{(\dfrac{dr}{d\varphi})}{\sqrt{(\dfrac{dr}{d\varphi})^{2} + (\dfrac{g_{\varphi\varphi}}{g_{rr}})}},
\end{align}
or more conveniently, the exact angle between the radial direction and the spatial direction of the light ray
is now given by
\begin{align} \label{2}
\tan(\psi) = \dfrac{\sqrt{\dfrac{g_{\varphi\varphi}}{g_{rr}}}}{\mid \dfrac{dr}{d\varphi} \mid}.
\end{align}
Thus, according to Eq.(\ref{2}), we have
\begin{align} 
\tan (\psi)_{GM} = \sqrt{\dfrac{\Big(1-\dfrac{2M}{r(\varphi)} \Big)  r_{p}\Big(r_{p} - \dfrac{Q^{2}}{M} \Big)}{\Big(1 - \dfrac{2M}{r_{p}}\Big) r(\varphi)\Big(r(\varphi) - \dfrac{Q^{2}}{M} \Big) - \Big(1 - \dfrac{2M}{r(\varphi)}\Big) r_{p}\Big(r_{p} - \dfrac{Q^{2}}{M} \Big)}},
\end{align}
for GMGHS spacetime,
\begin{align} 
\tan (\psi)_{Mag} = \sqrt{\dfrac{r_{p}^{4}\Big(1 - \dfrac{Q^{2}}{Mr_{p}}\Big)^{2}}{\dfrac{r_{p}^{2} \Big(1 - \dfrac{2M}{r_{p}}\Big) \Big(1 - \dfrac{Q^{2}}{Mr_{p}}\Big) r(\varphi)^{4} \Big(1 - \dfrac{Q^{2}}{Mr(\varphi)}\Big)^{2}  }{\Big(r(\varphi)^{2}(1 - \dfrac{2M}{r(\varphi)})(1-\dfrac{Q^{2}}{Mr(\varphi)})\Big)} - r_{p}^{4}\Big(1 - \dfrac{Q^{2}}{Mr_{p}}\Big)^{2}}},
\end{align}
for magnetically charged GMGHS spacetime, and
\begin{align} 
\tan (\psi)_{El} = \sqrt{\dfrac{r(\varphi)^{2} \Big(1+\dfrac{Q^{2} - 2M^{2}}{Mr(\varphi)}\Big) r_{p}^{2}\Big(r_{p}+\dfrac{Q^{2}}{M}\Big)^{2}}{r_{p}^{2} \Big(1+\dfrac{Q^{2}-2M^{2}}{Mr_{p}}\Big)-r(\varphi)^{2} \Big(1+\dfrac{Q^{2}-2M^{2}}{Mr(\varphi)}\Big) r_{p}^{2}\Big(r_{p}+\dfrac{Q^{2}}{M}\Big)^{2}}},
\end{align}
for electrically charged GMGHS spacetime.

where, $ r(\varphi) $ is the solutions of Eqs.(\ref{dr/dfi- GM})-(\ref{dr/dfi- El}) which solved in Eq.(\ref{rtfi}) for $ \epsilon=0 $. These now are valid for all light rays, not only for those rays
showing a small deflection as discussed in Refs.~\cite{Rindler:2007zz,Bhattacharya:2009rv,Bhattacharya:2010xh}.

\subsubsection{\textbf{Periastron advance of bound timelike orbits}}

In the case that $ R(r) $ \big(Eqs.(\ref{dr/dfi- GM})-(\ref{dr/dfi- El})\big) has at least two positive zeros, 
we may have a bound orbit for some initial values. 
The periastron advance $ \Delta_{peri} $ for such a bound orbit is given by
the difference of the $ 2\pi $-periodicity of the angle $ \varphi $ and the periodicity of the solution
$ r(\varphi) $ ~\cite{Kraniotis:2003ig,Kraniotis:2004cz},
\begin{equation}
\Delta^{Dilaton}_{peri}=2(\omega-\pi).
\end{equation}
The equations of motion for the static dilaton (GMGHS, magnetically GMGHS and  electrically GMGHS) 
black hole and the rotating dilaton (The Ker-Sen Dilaton-Axion) black hole, which is described in sec.\ref{field}, 
are polynomials of degree four and can be solved in terms of Weierstrass elliptic functions similarly.
So, for example, we consider the electrically GMGHS black hole and calculate physical data i.e. 
 the aphel $ r_{A} $ , the perihel $ r_{p} $, and the perihelion advance. 
 For the calculation of the perihelion precession and the orbital characteristics
of Mercury we use the following values for the physical constants:
\begin{equation}
c=2.99 792 458 \times10^{10} \rm{cm} s^{-1},\qquad \alpha_{S}=\dfrac{2GM_{\bigodot}}{c^{2}}= 2.953 250 08 \times10^{5} \rm{cm} .
\end{equation}
As free parameters we may use $\mathcal{L}$ and $E$ given in Refs.~\cite{Hackmann:2008zz,Kraniotis:2003ig},
\begin{equation}
\mathcal{L}=1.1849627128268641 \times 10^{-28} \rm{cm}^{-2} s^2,\qquad E=0.029979245417779875 \times10^{12} \rm{cm} s^{-1}.
\end{equation}
The two half-periods $\omega$ and $\omega^{\prime}$ are given by the following Abelian integrals
\begin{equation}
\omega=\int_{e_1}^{\infty}\frac{dt}{\sqrt{4t^3-g_2t-g_3}},\qquad \omega^{\prime}=\int_{-\infty}^{e_3}\frac{dt}{\sqrt{-4t^3+g_2t+g_3}}.
\end{equation}
The results are shown in table \ref{tab:peri}. Here, we used the rotation period 87.97 days of Mercury
and 100 SI-years per century to determine the unit $ arcsec$  $ cy^{-1}$ . 
It can be seen from table \ref{tab:peri}, that, by increasing the value of charge, 
the values of $\Delta_{peri}^{Dilaton}$ and $r_P$, decrease, but the value of $r_A$, increases. 
Moreover, the results all and particular the case of $ Q=0 $,  compare well to the 
results in Ref.~\cite{Hackmann:2008zz,Kraniotis:2003ig} and 
also to observations \cite{history.nasa}. 

%%%%%%%%%%%%%%

\begin{table}[h]
\begin{center}
\begin{tabular}{|l|l|l|l|}
%{|lccll|}
\hline
& $Q=0$  & $Q=\frac{\alpha_{S}}{17}$  & $Q=\frac{\alpha_{S}}{9}$  \\
\hline\hline

roots & $e_1=0.16666664004188$  & $e_1=0.16666664004230$   & $e_1=0.16666664004729$                         \\
 & $e_2=-0.08333331728350096$ & $e_2=-0.08333331688784412$  & $e_2=-0.08333331607958446$                          \\
&$e_3=-0.08333332275837917$ & $e_3=-0.08333332315446105$  & $e_3=-0.08333332396770639$                       
\\ \hline
$\omega$ & 3.1415929045225 & 3.1415929045185
  & 3.1415929044716
\\ \hline
$\omega^{\prime}$  & 18.660760760808i & 18.948164087533i
& 18.655684048383i
\\ \hline
$\tau$  & 6.0313874723363972i & 6.0133874723488761i
& 5.9382881919023788i
\\ \hline
$\Delta_{peri}^{Dilaton}$  & 42.97971108675784 & 42.97859778044177
& 42.970564827479413
\\ \hline
$r_P$  & $4.5970712924677\times 10^{12}$ & $4.4584998737211\times 10^{12}$
& $4.1747947337211\times 10^{12}$
\\ \hline
%\\ \hline
$r_A$ & $6.9817100360998 \times 10^{12}$ & $7.2034864923737 \times 10^{12}$

  & $7.6930132612827 \times 10^{12}$

\\ \hline\hline
\end{tabular}
\caption{Predictions for $\Delta_{peri}^{Dilaton}$, $r_P$, $r_A$ for the indicated choice for 
$\mathcal{L}=1.1849627128268641 \times 10^{-28} \rm{cm}^{-2} s^2$ and  
$E=0.029979245417779875 \times10^{12} \rm{cm} s^{-1}$.}
\label{tab:peri}
\end{center}
\end{table}

\clearpage

%%%%%%%%%%%%%%%

\section{Rotating Dilaton Black hole}\label{Rotating Dilaton}
In this section, we will discuss the geodesics in the rotating (The Ker-Sen Dilaton-Axion) dilaton black hole and present analytical
solutions of the equations of motion.

\subsection{Metric}\label{field}
In 1992, Sen \cite{Sen:1992ua} was able to find a charged, stationary, axially-symmetric solution \cite{Yazadjiev:1999ce} of
the field equations by using target space duality, applied to the classical Ker solution. The
line element of this solution can be written, in generalized Boyer–Linquist coordinates, as
\begin{align}
ds^{2}=-\big(1-\dfrac{2Mr}{\rho^{2}}\big)dt^{2}
+\dfrac{\rho^{2}}{\Delta}dr^{2}+\rho^{2}d\theta^{2} - \dfrac{4Mra\sin^{2}\theta}{\rho^{2}}dtd\varphi \nonumber\\
+\big(r(r+r_{\alpha})+a^{2}+\dfrac{2Mra^{2}\sin^{2}\theta}{\rho^{2}}\big)\sin^{2}\theta d\varphi^{2},
\end{align}
where
\begin{align}
\Delta_{r}=r(r+r_{\alpha})-2Mr+a^{2},
\end{align}
\begin{align}
\rho^{2}=r(r+r_{\alpha})+a^{2}\cos^{2}\theta .
\end{align}
Here $M$ is the mass of the black hole, $a$ is the angular momentum per unit mass of the black hole
$(= J/M)$ and $r_{\alpha} = \dfrac{Q^{2}}{M}$, where Q is the charge of the black hole. 
For $a = 0$, the Kerr-Sen black hole reduces to the Gibbons-Maeda-Garfinkle-Horowitz-Strominger (GMGHS) black hole 
and for $r_{\alpha} =0$, we get Kerr black hole. Further if both $a = 0$ and $r_{\alpha} =0$ then it reduces to Schwarzschild black hole.

\subsection{The geodesic equations}\label{geodesic}

In this section we discuss about geodesic equation and introduce effective potential and types of motion.

The Hamilton--Jacobi equation
\begin{equation}\label{Hamilton}
\dfrac{\partial S}{\partial\tau} +\frac{1}{2} \ g^{ij}\dfrac{\partial S}{\partial x^{i}}\dfrac{\partial S}{\partial x^{j}}=0, 
\end{equation}
can be solved with an ansatz for the action
\begin{equation}
\label{S}
S=\frac{1}{2}\varepsilon \tau - Et+L_{z}\phi +S_{\theta}(\theta) + S_{r} (r).
\end{equation}
The constants of motion are the energy $E$ and the angular momentum $L$  which are given
by the generalized momenta $P_{t}$ and $P_{\phi}$
\begin{equation}\label{constants of motion}
P_{t}=g_{tt}\dot{t}+g_{t \varphi}\dot{\varphi}=-E,  \qquad P_{\phi}=g_{\varphi \varphi}\dot\varphi +g_{t \varphi}\dot{t} =L.
\end{equation}
Using Eqs.(\ref{Hamilton})--(\ref{constants of motion}), we get
\begin{align}\label{ds/dr.ds/dtheta}
-\Delta_{r}\big(\dfrac{ds}{dr}\big)^{2} - \varepsilon r^{2}+\dfrac{1}{\Delta_{r}}\big((a^{2}+r(r+r_{\alpha}))E - aL\big)^{2} - \big(aE - L\big)^{2} = \nonumber\\
\big(\dfrac{ds}{d\theta}\big)^{2}+\varepsilon a^{2}\cos^{2}\theta + \big(\dfrac{L^{2}}{\sin^{2}\theta} - a^{2}E^{2}\big)\cos^{2}\theta,
\end{align}
where each side depends on $r$ or $\theta$ only. 
From the separation ansatz Eq.(\ref{S}) and with the help of the Carter \cite{K} constant,
we derive the equations of motion:
\begin{align}\label{Joda}
\rho^{4}(\dfrac{dr}{d\tau})^{2}=-\Delta_{r}(K+\varepsilon r(r+r_{\alpha}))+\big((a^{2}+r(r+r_{\alpha}))E-aL \big)^{2}=R(r),
\end{align}
\begin{align}\label{thetatho}
\rho^{4}(\dfrac{d\theta}{d\tau})^{2}=\Delta_{\theta}(K-\varepsilon a^{2}\cos^{2}\theta)-\dfrac{1}{\sin^{2}\theta}\big(aE \sin^{2}\theta -L \big)^{2} =\Theta(\theta),
\end{align}
\begin{align}
\rho^{2}(\dfrac{d\varphi}{d\tau})=\dfrac{a}{\Delta_{r}}\big((a^{2}+r(r+r_{\alpha})) E - aL \big)-\dfrac{1}{\sin^{2}\theta}(a E \sin^{2}\theta - L),
\end{align}
\begin{align}\label{ttho}
\rho^{2}(\dfrac{dt}{d\tau})=\dfrac{(a^{2}+r(r+r_{\alpha}))}{\Delta_{r}}\big((a^{2}+r(r+r_{\alpha})) E - aL \big) - a (a E \sin^{2}\theta - L).
\end{align}
In the following, we will explicitly solve these equations. 
Eq.(\ref{Joda}) suggests the introduction of an effective potential $V_{eff,r}$ such that $V_{eff,r}=E$ 
corresponds to $(\dfrac{dr}{d\tau})^{2}=0$, 
\begin{equation}
V_{eff,r}=\dfrac{L a\pm \sqrt{\Delta_{r}(K+\varepsilon r(r+r_{\alpha}))}}{a^{2}+r(r+r_{\alpha})},
\end{equation} 
where $(\dfrac{dr}{d\tau})^{2}\geq 0$ for $E\leq V_{eff,r}^{-}$ and $E\geq V_{eff,r}^{+}$. 
In the same way an effective potential
corresponding to Eq.(\ref{thetatho}) can be introduced
\begin{align}
V_{eff,\theta}=\dfrac{L \pm \sqrt{\sin^{2}\theta (K-\varepsilon a^{2}\cos^{2}\theta)}}{a \sin^{2}\theta},
\end{align}
but here $(\dfrac{d\theta}{d\tau})^{2}\geq 0$ for $V_{eff,\theta}^{-}\leq E \leq V_{eff,\theta}^{+}$.
\\
Introducing the Mino time $\lambda$ \cite{Y.Mino} connected to the proper time 
$\tau$ by $\dfrac{d\tau}{d\lambda}=\rho^{2}$, the equations of motions read
\begin{align}\label{d}
(\dfrac{dr}{d\lambda})^{2}=-\Delta_{r}(K+\varepsilon r(r+r_{\alpha}))+\big((a^{2}+r(r+r_{\alpha}))E-aL \big)^{2}=R(r),
\end{align}
\begin{align}
(\dfrac{d\theta}{d\lambda})^{2}=\Delta_{\theta}(K-\varepsilon a^{2}\cos^{2}\theta)-\dfrac{1}{\sin^{2}\theta}\big(aE \sin^{2}\theta -L \big)^{2} =\Theta(\theta),
\end{align}
\begin{align}
(\dfrac{d\varphi}{d\lambda})=\dfrac{a}{\Delta_{r}}\big((a^{2}+r(r+r_{\alpha})) E - aL \big)-\dfrac{1}{\sin^{2}\theta}(a E \sin^{2}\theta - L),
\end{align}
\begin{align}\label{t}
(\dfrac{dt}{d\lambda})=\dfrac{(a^{2}+r(r+r_{\alpha}))}{\Delta_{r}}\big((a^{2}+r(r+r_{\alpha})) E - aL \big) - a (a E \sin^{2}\theta - L).
\end{align}
 We introduce dimensionless quantities for rescale the parameters
\begin{align}
\tilde{r}=\dfrac{r}{M} ,\quad  \tilde{a}=\dfrac{a}{M} ,\quad \tilde{t}=\dfrac{t}{M} , \quad  \tilde{L}=\dfrac{L}{M} , \quad  
\tilde{K}=\dfrac{K}{M^{2}}, \quad  \tilde{r}_{\alpha}=\dfrac{r_{\alpha}}{M}, \quad  \gamma = M\lambda.
\end{align}
Then the equations (\ref{d}) - (\ref{t}), can be rewritten as
\begin{align}\label{drd}
\left(\dfrac{d\tilde{r}}{d\gamma}\right)^{2}=-\Delta_{\tilde{r}}(\tilde{K}+\varepsilon
\tilde{r}(\tilde{r}+\tilde{r}_{\alpha}))+ \left( (\tilde{a}^{2}+\tilde{r}(\tilde{r}+\tilde{r}_{\alpha}))E
-\tilde{a}\tilde{L} \right)^{2}=\tilde{R}(\tilde{r}),
\end{align}
\begin{align}\label{dthetad}
\left(\dfrac{d\theta}{d\gamma}\right)^{2}=(\tilde{K}
-\varepsilon \tilde{a}^{2}\cos^{2}\theta)
-\dfrac{1}{\sin^{2}\theta}\left(\tilde{a}E \sin^{2}\theta
-\tilde{L} \right)^{2}=\tilde{\Theta}(\theta),
\end{align}
\begin{align}\label{dphi}
\left(\dfrac{d\varphi}{d\gamma}\right)=\dfrac{\tilde{a}}{{\Delta_{\tilde{r}}}}
\left( (\tilde{a}^{2}+\tilde{r}(\tilde{r}+\tilde{r}_{\alpha}))E
-\tilde{a}\tilde{L} \right)
-\dfrac{1}{\sin^{2}\theta} \big(\tilde{a}E
\sin^{2}\theta -\tilde{L}  \big),
\end{align}
\begin{align}\label{dtd}
\left(\dfrac{d\tilde{t}}{d\gamma}\right)=\dfrac{(\tilde{a}^{2}+\tilde{r}(\tilde{r}+\tilde{r}_{\alpha})}{{\Delta_{\tilde{r}}}}
\left( (\tilde{a}^{2}+\tilde{r}(\tilde{r}+\tilde{r}_{\alpha}))E
-\tilde{a}\tilde{L} \right)
-a\big(\tilde{a}E
\sin^{2}\theta -\tilde{L}\big).
\end{align}

 \subsubsection{\textbf{Types of latitudinal motion}}
 
 In this subsection and next subsection, we use of function $ \tilde{\Theta}(\theta) $ in equation 
 (\ref{dthetad}) and polynomial $ \tilde{R}(\tilde{r}) $ in equation (\ref{drd}), 
 to determine the possible orbits of light and test particles.

\begin{figure}[h]
 \centering
 \includegraphics[width=0.4\textwidth]{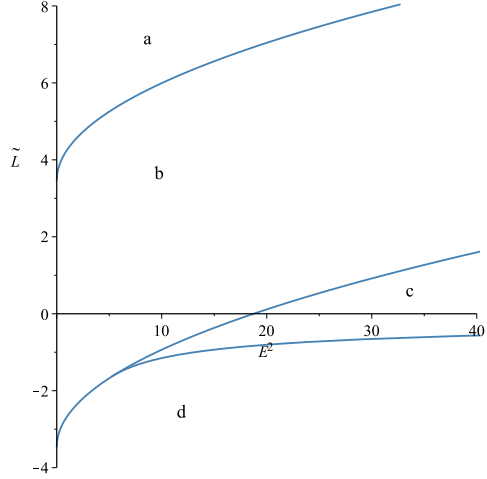}
 \caption{ \label{fig:parameterplot-theta} $\epsilon =1$, $\tilde{a}=0.8$,  $\tilde{K}=12$,: Parametric $\tilde{L}$-$E^2$-diagram for the function  $\tilde{\Theta}$. $\tilde{\Theta}$ possesses one zero in region b and two zeros in region c. In the a and d areas geodesic motion is not possible.}
\end{figure}

First we substitute $\upsilon = cos^{2}\theta$ with $ \theta \in [0, 1]$ in the function $\tilde{\Theta}(\theta)$:
 \begin{align}
 \tilde{\Theta}(\upsilon)=\tilde{K} - \varepsilon\tilde{a}^{2}\upsilon - \big(\tilde{a}^{2}E^{2}(1-\upsilon)-2\tilde{L} \tilde{a}E+\dfrac{\tilde{L}^{2}}{(1-\upsilon)} \big) .
 \end{align} 
 Geodesic motion is possible if $\tilde{\Theta}(\theta)\geqslant 0$, then real values of the coordinate $ \theta $ is obtained. 
 The number of zeros only changes if a zero crosses 
 $0$ or $1$, or if a double zero occurs. $\upsilon=0$, is a zero of $\tilde{\Theta}$, if
 \begin{align}
 \tilde{\Theta}(\upsilon =0)=\tilde{K}- \big(\tilde{a}^{2}E^{2}-2\tilde{L}\tilde{a}E+\tilde{L}^{2} \big) ,
 \end{align}
 and therefore
 \begin{align}
 \tilde{L}=\tilde{a}E\pm \sqrt{\tilde{K}}.
 \end{align}
 Since $\upsilon=1$, is a pole of $\tilde{\Theta}(\upsilon)$ for $\tilde{L}\neq 0$, it is only possible that $\upsilon=1$, is a zero of $\tilde{\Theta}(\upsilon)$, if $\tilde{L}=0$,
 \begin{align}
 \tilde{\Theta}(\upsilon =1, \tilde{L}=0)=\tilde{K}-\varepsilon \tilde{a}^{2}.
 \end{align}
To remove the pole of $\tilde{\Theta}(\upsilon)$ at $\upsilon = 1$, we consider
 \begin{align}\label{thetanoo}
 \tilde{\Theta}^{'}({\upsilon})=(1-\upsilon)(\tilde{K}-\varepsilon \tilde{a}^{2}\upsilon)-\big( \tilde{a}E(1-\upsilon)-\tilde{L} \big)^{2},
 \end{align}
where, $ \tilde{\Theta}({\upsilon})  =\frac{1}{1-\upsilon}\tilde{\Theta}^{'}({\upsilon}) $. Then double zeros fulfill the conditions
 \begin{align}\label{x}
  \tilde{\Theta}^{'}({\upsilon})=0, \qquad  \text{and}  \qquad \frac{ d\tilde{\Theta}^{'}({\upsilon})}{d\upsilon}=0.
 \end{align}
From the condition of $\upsilon=0$, and both conditions of Eqs.~(\ref{x}), we can plot parametric $\tilde{L}$-$E^2$-diagrams for types of latitudinal motion (see Fig.~\ref{fig:parameterplot-theta}). These reveal two regions in which geodesic motion is possible. 

\subsubsection{\textbf{Types of radial motion}}
\begin{figure}[h]
 \centering
 \includegraphics[width=0.4\textwidth]{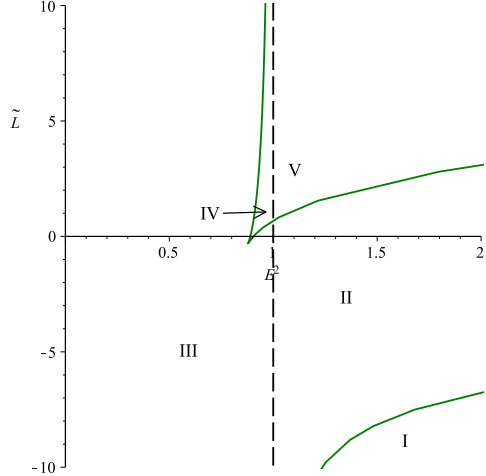}
 \caption{ \label{pic:parametric-diagrams1} $\epsilon =1$, $\tilde{a}=0.8$,  $\tilde{K}=12$,: Parametric $\tilde{L}$-$E^2$-diagram 
 of the $\tilde{r}$-motion. The polynomial $\tilde{R}$ has 0 zero in region I, 2 zeros in region II, 2 positive zeros in region III, 4 positive zeros in region IV, 3 positive and 1 negative zeros in region V. }
\end{figure}

The zeros of the polynomial $\tilde{R}$, are the turning points of orbits of
light and test particles and therefore $\tilde{R}$, determines the possible
types of orbits,
\begin{align}
\tilde{R}(\tilde{r})=-\Delta_{\tilde{r}}(\varepsilon \tilde{r}(\tilde{r}+\tilde{r}_{\alpha})
+\tilde{K})+ \big((\tilde{a}^{2}
+\tilde{r}(\tilde{r}+\tilde{r}_{\alpha}))E-\tilde{a}\tilde{L}\big)^{2}.
\end{align}
The type of an orbit is determined by the number of real zeros of the polynomial $\tilde{R}$. This number changes if a double zero occurs,
\begin{equation}
	\tilde{R}(\tilde{r})=0, \qquad \text{and} \qquad \frac{d \tilde{R}(\tilde{r})}{d\tilde{r}}=0 .
	\label{eqn:doublezero}
\end{equation}
Taking both these conditions into account, we can plot parametric $\tilde{L}$-$E^2$-diagrams 
which show five regions with different numbers of zeros, (see Fig.~\ref{pic:parametric-diagrams1}). If null geodesics
($ \epsilon=0 $), are considered, the regions III and IV, vanish from the parametric $\tilde{L}$-$E^2$-diagram.

%\clearpage

\subsection{Analytical solution of geodesic equations}\label{analytical solutions}
In this section, the analytical solutions of the geodesic equations (\ref{drd})--(\ref{dtd}), in 
Rotating Kerr-Sen Dilaton-Axion black hole spacetime, are presented 
in the terms of the elliptic Weierstrass $ \wp $, $\zeta$ and $\sigma$ functions. 
Each equation will be treated separately.

\subsubsection{\textbf{r motion}}\label{rmotion}
The differential equation that describes the dynamics of $r$ Eq.(\ref{drd})
\begin{align}
\left(\dfrac{d\tilde{r}}{d\gamma}\right)^{2}=-\Delta_{\tilde{r}}(\tilde{K}+\varepsilon
\tilde{r}(\tilde{r}+\tilde{r}_{\alpha}))+ \left( (\tilde{a}^{2}+\tilde{r}(\tilde{r}+\tilde{r}_{\alpha}))E
-\tilde{a}\tilde{L} \right)^{2}=\tilde{R}(\tilde{r}),
\end{align}
is a polynomial of degree four for both $\varepsilon=1$ and $\varepsilon=0$ and, therefore, 
it is of elliptic type if $\tilde{R}(\tilde{r})$ has only simple zeros.
With the substitutions $\tilde{r}=u^{-1}+\tilde{r}_{\tilde{R}}$ ,
where $\tilde{r}_{\tilde{R}}$ is a zero of $\tilde{R}$, transforms the problem to
\begin{equation}\label{du/dfi}
(\dfrac{du}{d\varphi})^{2}= R_{3}(u) = \sum_{j=1}^{3} a_{j} u^{j} ,\qquad u(\varphi_{0})=u_{0} ,
\end{equation}
with a polynomial $R_{3}$ of degree 3. Where
\begin{equation}
a_{j}=\dfrac{1}{(4-j)!}\dfrac{d^{(4-j)}R}{d \tilde{r}^{(4-j)}} (\tilde{r}_{R}) .
\end{equation} 
Afterward, the substitution $u=\frac{1}{a_{3}}(4y-\frac{a_{2}}{3})$, implies
\begin{align}\label{dy}
(\dfrac{dy}{d\gamma})^{2}=4y^{3}-g_{2}y-g_{3},
\end{align}
where
\begin{align}
g_{2}=\frac{1}{16} \big( \frac{4}{3} a_{2}^{2}-4a_{1}a_{3} \big),
\end{align}
\begin{align}\label{g3}
g_{3}=\frac{1}{16} \big( \frac{1}{3}a_{1}a_{2}a_{3}
-\frac{2}{27}a_{2}^{3}-a_{0}a_{3}^{2} \big),
\end{align}
are the Weierstrass invariants. the differential equation (\ref{dy}),
is elliptic of first kind, which can be solved by 
\begin{align}\label{PW}
y(\gamma)=\wp \big(\gamma -\gamma_{\theta , in};g_{2},g_{3} \big).
\end{align}
Accordingly, the solution of Eq.(\ref{dthetad}), is given by
\begin{align}\label{rr}
\tilde{r}(\gamma)=\dfrac{a_{3}}{4\wp(\gamma
-\gamma_{\tilde{r},in};g_{2},g_{3})-\frac{a_{2}}{3}}+\tilde{r}_{\tilde{R}},
\end{align}
where,
$\gamma_{\tilde{r},in}=\gamma_{0}+\int_{y_{0}}^{\infty}\dfrac{dy'}
{\sqrt{4y'^{3}-g_{2}y'-g_{3}}}$ , with
$y_{0}=\frac{a_{3}}{4(\tilde{r}_{0}-r_{R})}+\frac{a_{2}}{12}$ ,
depends only on the initial values $\gamma_{0}$ and $\tilde{r}_{0}$ .

\subsubsection{\textbf{$\theta$ motion}}\label{thetam}
The differential equation (\ref{dthetad}),
\begin{align}
(\dfrac{d\theta}{d\gamma})^{2}=
\tilde{\Theta}(\theta)=(\tilde{K}-\varepsilon
\tilde{a}^{2}\cos^{2}\theta)-\dfrac{1}{\sin^{2}\theta} \big(
\tilde{a}E \sin^{2}\theta -\tilde{L}\Xi \big)^{2},
\end{align}
which can be simplified by the substitution $\upsilon=cos^{2}\theta$, yielding
\begin{align}\label{doo}
(\dfrac{d\upsilon}{d\gamma})^{2}=4\upsilon \tilde{\Theta}^{'}({\upsilon})=4\upsilon(1-\upsilon)(\tilde{K}-\varepsilon \tilde{a}^{2}\upsilon)-4\upsilon\big( \tilde{a}E(1-\upsilon)-\tilde{L}\Xi \big)^{2} .
\end{align}
where, $ \tilde{\Theta}^{'}({\upsilon})=(1-\upsilon)\tilde{\Theta}({\upsilon}) $ . 
The differential equation (\ref{doo}), for both $ \varepsilon = 1 $ and $ \varepsilon = 0 $, is a polynomial of
degree three and with assume that $\tilde{\Theta}^{'}({\upsilon})$
has only simple zeros, Then it can be solved in terms of the
Weierstrass $\wp$ elliptic function. Thus, with the standard substitution
$\upsilon =\frac{1}{b_{3}} \big(4y-\frac{b_{2}}{3} \big)$, where
$4\upsilon \tilde{\Theta}^{'}({\upsilon})=\sum_{i=1}^3 b_{i}\upsilon^{i}$,
transforms the problem to the form Eq.(\ref{dy}). The solution is
then given by
\begin{align}
\theta \big( \gamma \big)=\arccos \Big( \pm \sqrt{\frac{4}{b_{3}}\wp
(\gamma - \gamma_{\theta ,in}; g_{2},g_{3})-\frac{b_{2}}{3 b_{3}}}
\Big) , 
\end{align}
where, $\gamma_{\theta ,in}=\gamma_{0}+
\int_{y_{0}}^{\infty}\dfrac{dy'} {\sqrt{4y'^{3}-g_{2}y'-g_{3}}}$ ,
with $y_{0}=\dfrac{b_{3}}{4 \cos^{2}(\theta_{0})}+\dfrac{b_{2}}{12}$ ,
depends on the initial values $\gamma_{0}$ and $\theta_{0}$ only.
The sign of the square root depends on whether $\theta(\gamma)$,
should be in $(0,\frac{\pi}{2})$ (positive sign) or in
$(\frac{\pi}{2},\pi)$ (negative sign), and reflects the symmetry of
the $\theta$ motion with respect to the equatorial plane
$\theta=\frac{\pi}{2}$.

\subsubsection{\textbf{$\varphi$ motion}}\label{fii}
In this part and next part, we solve equation of $\varphi$ and $ t $ motion according to Weierstrass $ \wp $, $\zeta$ and $\sigma$ functions.
For analysis $\varphi$ motion, we use Eq.(\ref{dphi}),
\begin{align}
(\dfrac{d\varphi}{d\gamma})=\dfrac{E(\tilde{a}^{3}+\tilde{r}(\tilde{r}+\tilde{r}_{\alpha})\tilde{a})-\tilde{L} \tilde{a}^{2}}{\Delta_{\tilde{r}}}-\dfrac{1}{\sin^{2}\theta} \big( \tilde{a} E \sin^{2}\theta -\tilde{L} \big).
\end{align}
This equation can be splitted in a part dependent only on $\tilde{r}$ and in a part only dependent on $\theta$. Integration yields
\begin{align}\label{phi}
\varphi - \varphi_{0}= \bigg[ \int_{\gamma_{0}}^{\gamma} \dfrac{E(\tilde{a}^{3}+\tilde{r}(\tilde{r}+\tilde{r}_{\alpha})\tilde{a})-\tilde{L} \tilde{a}^{2}}{\Delta_{\tilde{r}(\gamma)}}d\gamma - \int_{\gamma_{0}}^{\gamma}\dfrac{\tilde{a} E \sin^{2}\theta -\tilde{L}}{\sin^{2}\theta (\gamma)}d\gamma \bigg] \nonumber\\ = \bigg[ \int_{\tilde{r}_{0}}^{\tilde{r}} \dfrac{E(\tilde{a}^{3}+\tilde{r}^{2}\tilde{a})-\tilde{L} \tilde{a}^{2}}{\Delta_{\tilde{r}}\sqrt{\tilde{R}}}d\tilde{r} - \int_{\theta_{0}}^{\theta}\dfrac{\tilde{a} E \sin^{2}\theta -\tilde{L}}{\sin^{2}\theta \sqrt{\tilde{\Theta}(\theta)}}d\theta \bigg] \nonumber\\
=\bigg[ \tilde{I}_{r}-\tilde{I}_{\theta} \bigg]  ,
\end{align}
where, we substituted $\tilde{r}=\tilde{r}(\gamma)$, i.e. $\frac{d\tilde{r}}{d\gamma}=\sqrt{\tilde{R}}$, in the first and $\theta=\theta(\gamma)$, i.e. $\frac{d\theta}{d\gamma}=\sqrt{\tilde{\Theta}(\theta)}$, in the second integral. We will solve now the two integrals in Eq.(\ref{phi}) separately.
Let us consider the integral
\begin{align}
I_{\theta}=\int_{\theta_{0}}^{\theta}\dfrac{(\tilde{a} E \sin^{2}\theta -\tilde{L})d\theta}{\sin^{2}\theta \sqrt{\tilde{\Theta}(\theta)}},
\end{align}
which can be transformed to the simpler form
\begin{align}
I_{\theta}=\mp \int_{\upsilon_{0}}^{\upsilon}\dfrac{\tilde{a}E (1-\upsilon)-\tilde{L}}{(1-\upsilon)\sqrt{4\upsilon \tilde{\Theta}^{'}({\upsilon})}}d\upsilon ',
\end{align}
by the substitution $\upsilon=\cos^{2}\theta$, where $\tilde{\Theta}_{\upsilon}$, is defined in (\ref{thetanoo}). Here, we have to pay special attention to the integration path. If $\theta \in (0,\frac{\pi}{2}]$, we have $\cos \theta= + \sqrt{\upsilon}$ but for $\theta \in [\frac{\pi}{2},\pi)$, then $\cos \theta= - \sqrt{\upsilon}$. Accordingly, we first have to split the integration path from $\theta_{0}$ to $\theta$ such that every piece is fully contained in the interval $(0,\frac{\pi}{2}]$, or $[\frac{\pi}{2},\pi)$, and then to choose the appropriate sign of the square root of $\upsilon$. In the following we assume for simplicity that $\cos \theta= +\sqrt{\upsilon}$. If $4\upsilon\tilde{\Theta_{\upsilon}}$ has only simple zeros, $I_{\theta}$ is of elliptic type and of third kind. We transform analogously to section (\ref{thetam}), to the standard Weierstrass form by $\upsilon=\frac{1}{a_{3}}(4y-\frac{a_{2}}{3})$. Then, the solution to $I_{\theta}$, is given by
\begin{align}\label{thetafi}
I_{\theta}=-\dfrac{\mid b_{3} \mid}{b_{3}}\bigg\lbrace (\tilde{a}E)(\gamma - \gamma_{0}) + \sum_{i=1}^{4}\dfrac{b_{3}\tilde{L}}{4 \wp '(\upsilon_{i})} \bigg( \zeta(\upsilon_{i})(\gamma-\gamma_{0}) \nonumber\\+ \log \dfrac{\sigma (\gamma -\gamma_{i})}{\sigma(\gamma_{0}-\gamma_{i})} \bigg) \bigg\rbrace ,
\end{align}
where, the constants $b_{i}$, are defined as in section (\ref{thetam}), $\wp=\wp(\gamma)=\gamma - \gamma_{\theta,in}$, and $\gamma_{0}=\wp(\gamma_{0})$ \cite{Hackmann:2010zz}.

Now, we solve the $\tilde{r}$ dependent integral in (\ref{phi}),
\begin{align}
I_{r}= \int_{\tilde{r}_{0}}^{\tilde{r}} \dfrac{E(\tilde{a}^{3}+\tilde{r}^{2}\tilde{a})-\tilde{L} \tilde{a}^{2}}{\Delta_{\tilde{r}}\sqrt{\tilde{R}}}d\tilde{r}.
\end{align}
If we consider that $\tilde{R}$, has only simple zeros, $I_{r}$ is of elliptic type and third kind. We transform analogously to section (\ref{rmotion}), to the standard Weierstrass form by $\tilde{r}=\pm 1/u+\tilde{r}_{\tilde{R}}$, with a zero $\tilde{r}_{\tilde{R}}$ of $\tilde{R}$, and then $u=\frac{1}{a_{3}}(4y-\frac{a_{2}}{3})$. Afterward, we simplify the integrand by a partial fraction decomposition and this integral can now be solved as
\begin{align}\label{rfi}
I_{r}(\gamma)=-\dfrac{\mid a_{3} \mid}{a_{3}}\bigg\lbrace (\dfrac{E(\tilde{a}^{3}+ \tilde{r}_{\tilde{R}}( \tilde{r}_{\tilde{R}}+\tilde{r}_{\alpha})\tilde{a})-\tilde{L} \tilde{a}^{2}}{\Delta_{ \tilde{r}_{\tilde{R}}(\gamma)}})(\gamma - \gamma_{0}) + \sum_{i,j=1}^{2}\dfrac{C_{i}}{\wp '(\gamma_{ij})} \bigg( \zeta(\upsilon_{ij})(\gamma-\gamma_{0}) \nonumber\\+ \log \dfrac{\sigma (\gamma -\gamma_{ij})}{\sigma(\gamma_{0}-\gamma_{ij})} \bigg) \bigg\rbrace ,
\end{align}
where, $\wp(\gamma_{ij} - \gamma_{r,in})=y_{i}$, and $C_{i}$, are the coefficients of the partial fractions $(y-y_{i})^{-1}$.

\subsubsection{\textbf{t motion}}
The equation for $t$ Eq.(\ref{dtd}),
\begin{align}\label{dtd}
\left(\dfrac{d\tilde{t}}{d\gamma}\right)=\dfrac{(\tilde{a}^{2}+\tilde{r}(\tilde{r}+\tilde{r}_{\alpha})}{{\Delta_{\tilde{r}}}}
\left( (\tilde{a}^{2}+\tilde{r}(\tilde{r}+\tilde{r}_{\alpha}))E
-\tilde{a}\tilde{L} \right)
-a\big(\tilde{a}E\sin^{2}\theta -\tilde{L}\big),
\end{align}
has the same structure as the equation for the $\varphi$ motion. An integration yields
\begin{align}
\tilde{t}-\tilde{t}_{0}=\bigg[ \int_{\gamma_{0}}^{\gamma}{\dfrac{(\tilde{a}^{2}+\tilde{r}(\tilde{r}+\tilde{r}_{\alpha})}{{\Delta_{\tilde{r}}}}
\left( (\tilde{a}^{2}+\tilde{r}(\tilde{r}+\tilde{r}_{\alpha}))E
-\tilde{a}\tilde{L} \right)} d\gamma - \int_{\gamma_{0}}^{\gamma}{a\big(\tilde{a}E\sin^{2}\theta -\tilde{L}\big)} \nonumber\\
=\bigg[ \tilde{I}_{r}-\tilde{a}\tilde{I}_{\theta} \bigg]. 
\end{align}
Because we already demonstrated the solution procedure, we only give here the results for the most general cases. If $4\tilde{\Theta}_{\upsilon}$, where $\tilde{\Theta}_{\upsilon}$ is defined in Eq.(\ref{thetanoo}), has only simple zeros the solution of the $\theta$ dependent part is given by
\begin{align}
\tilde{I}_{\theta}= \dfrac{4\tilde{a}E}{b_{3}} \zeta(\gamma - \gamma_{0})+\dfrac{3\tilde{a}E b_{3}+\tilde{a}E b_{2}-3b_{3}\tilde{L} }{3b_{3}} (\gamma -\gamma_{0}).
\end{align}
And the solution of the $\tilde{r}$ dependent part is given by
\begin{align}\label{rfi}
I_{r}=C_{0}(\gamma - \gamma_{0}) + \sum_{i=1}^{3}\sum_{j=1}^{2}\dfrac{C_{i1}\tilde{L}}{\wp^{'}(\upsilon_{ij})} \bigg( \zeta(\upsilon_{ij})(\gamma-\gamma_{0}) + \log \dfrac{\sigma (\gamma -\gamma_{ij})}{\sigma(\gamma_{0}-\gamma_{ij})} \bigg) \nonumber\\
-\sum_{j=1}^{2}\dfrac{C_{32}\tilde{L}}{\wp^{'}(\upsilon_{3j})} \bigg( \dfrac{\wp(\upsilon_{3j})\wp '(\upsilon_{3j})
+2\wp^{''}(\upsilon_{3j})\zeta(\upsilon_{3j})}{\wp^{'}(\upsilon_{3j})^{2}}(\gamma-\gamma_{0}) \nonumber\\
+\zeta(\gamma -\gamma_{3j})+ 2\wp^{''}(\upsilon_{3j})\log \dfrac{\sigma (\gamma -\gamma_{3j})}{\sigma(\gamma_{0}-\gamma_{3j})} \bigg), 
\end{align}
where, $C_{0}$, is a constant and $C_{ik}$ are the coefficients of the partial fractions $(y-y_{i})^{-k}$.

\subsection{Orbits}

\begin{figure}[h]
	\centering
	\subfigure[]{
		\includegraphics[width=0.48\textwidth]{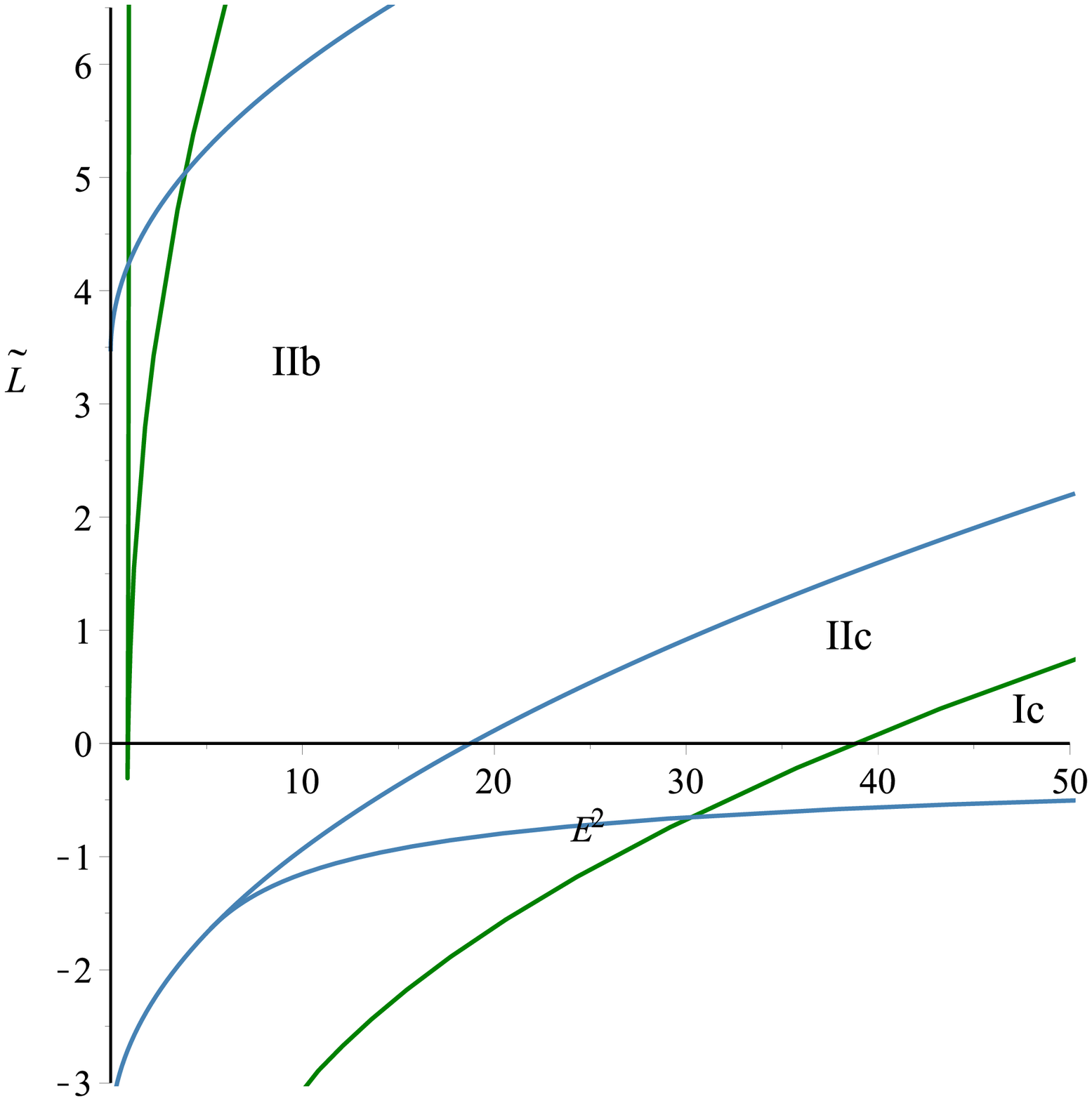}
	}
	\subfigure[closeup of figure (a)]{
		\includegraphics[width=0.48\textwidth]{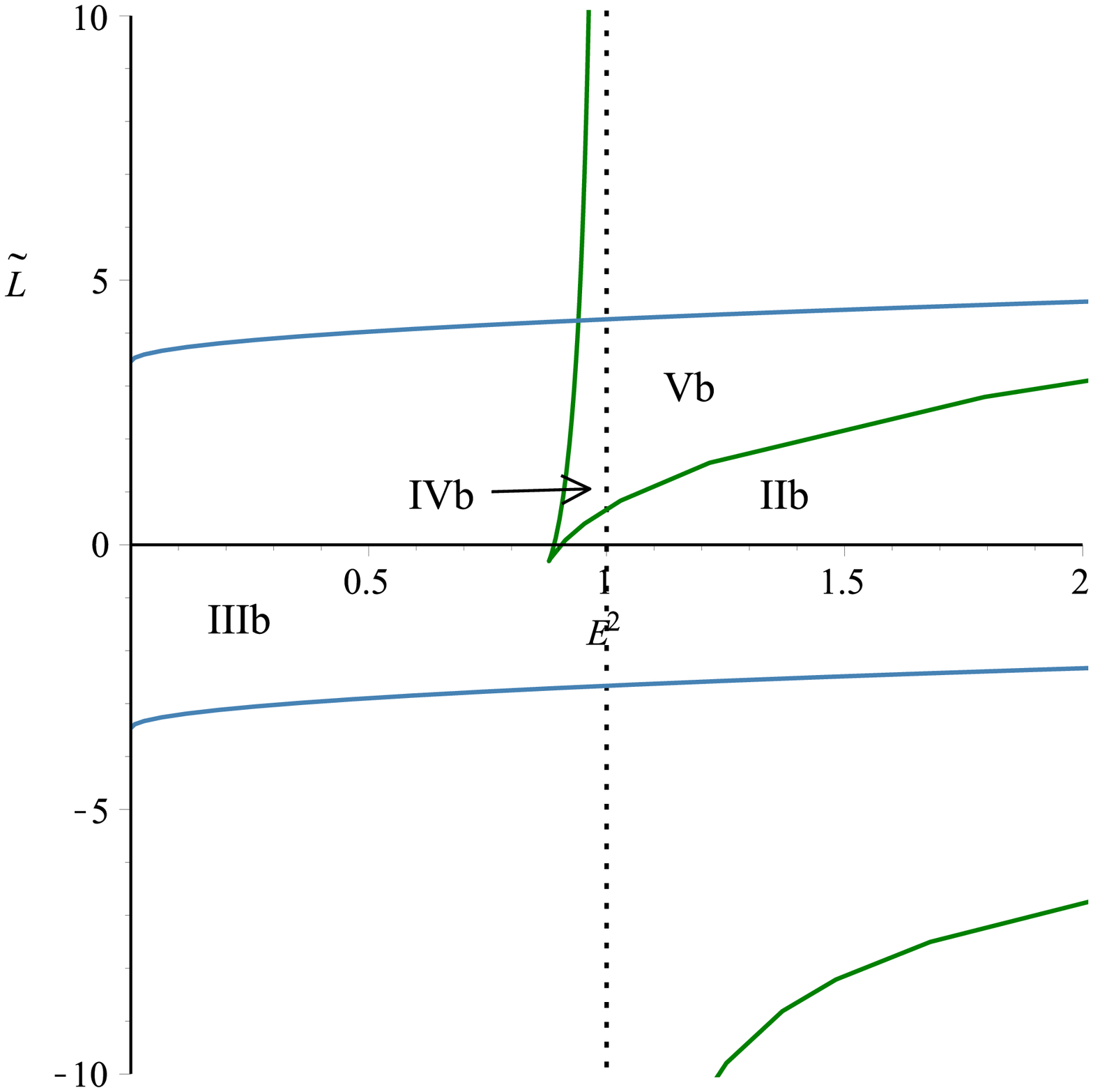}
	%}
	%\subfigure[$\epsilon=1$, $\tilde{a}=0.9$, $\tilde{K}=0.3$, $R_0=-4\cdot 10^{-5}$, $q=0.2$]{
	%	\includegraphics[width=.48\textwidth]{parameterplot3.eps}
	}
	\caption{\label{pic:parametric-diagrams} $\epsilon=1$, $\tilde{a}=0.8$, $\tilde{K}=12$. Combined $\tilde{L}$-$E^2$-diagrams of the $\tilde{r}$-motion (green lines) and $\theta$-motion (blue lines). In regions marked with the letter b, the orbits cross $\theta=\frac{\pi}{2}$, but not $\tilde{r}=0$. Whereas in regions marked with the letter c, $\tilde{r}=0$ can be crossed but $\theta=\frac{\pi}{2}$ is never crossed.}
\end{figure}

\begin{figure}[h]
	\centering
	\subfigure[]{
		\includegraphics[width=0.3\textwidth]{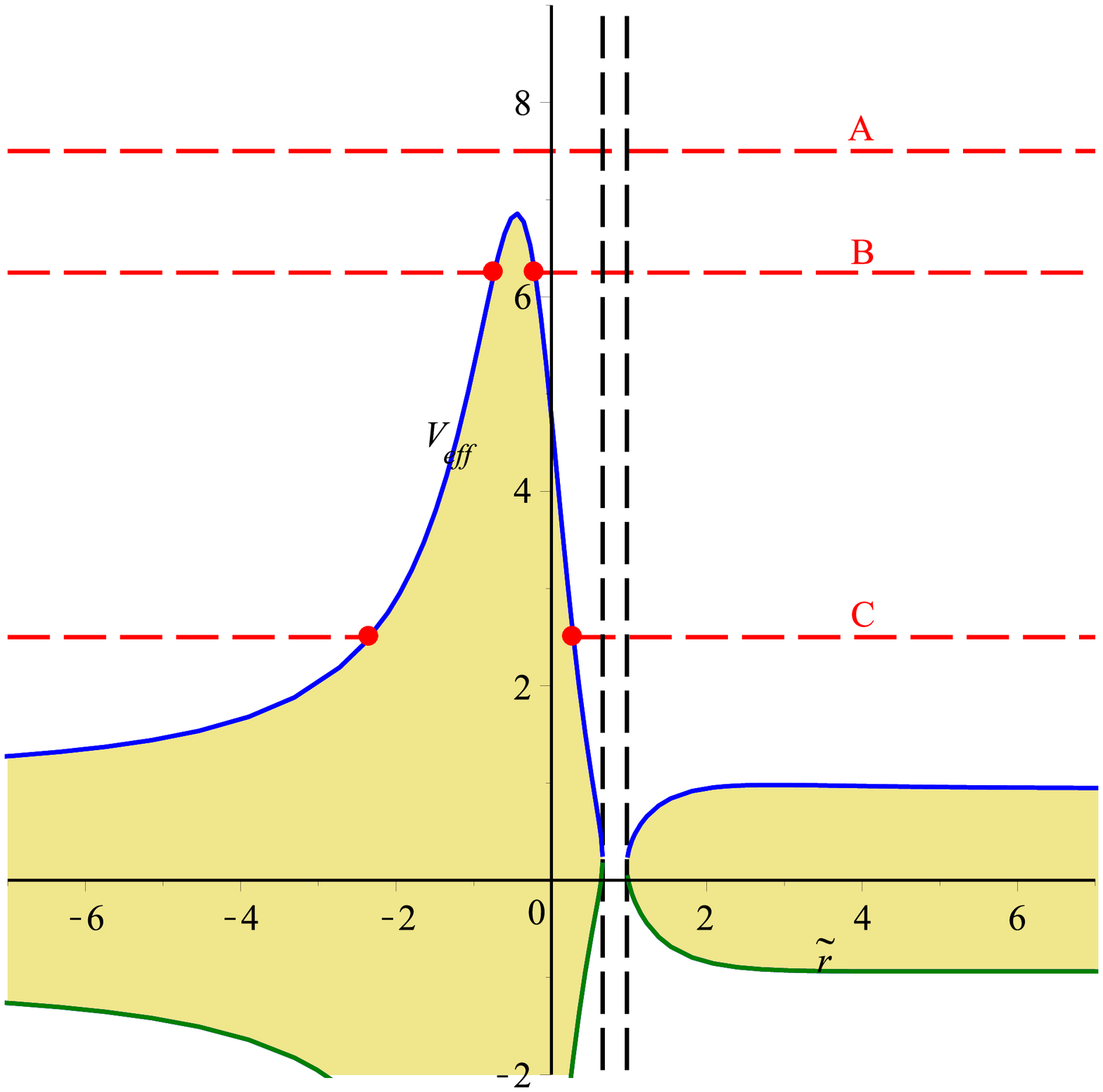}
	}
	\subfigure[]{
		\includegraphics[width=0.33\textwidth]{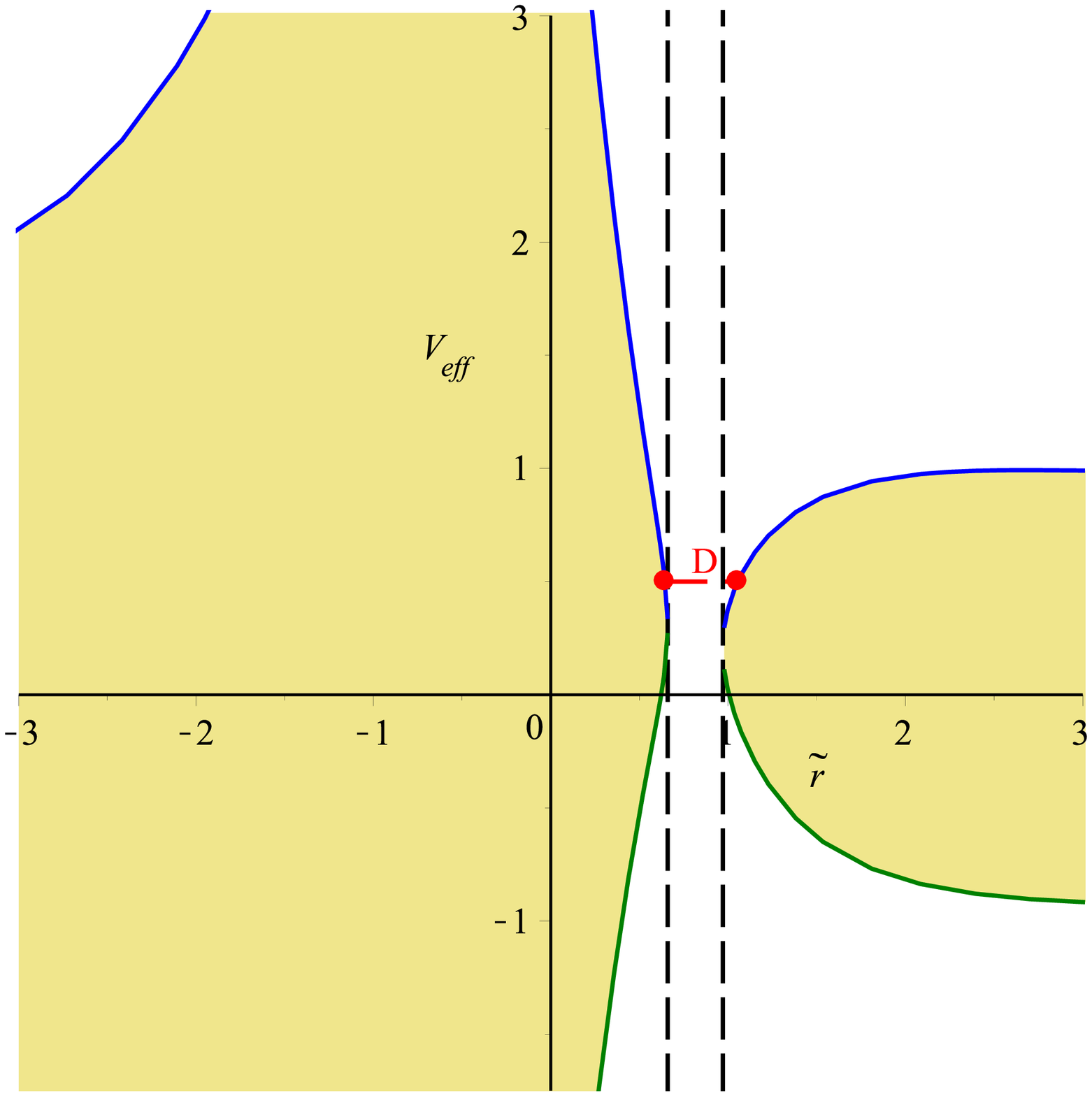}
	}
	\subfigure[]{
		\includegraphics[width=0.33\textwidth]{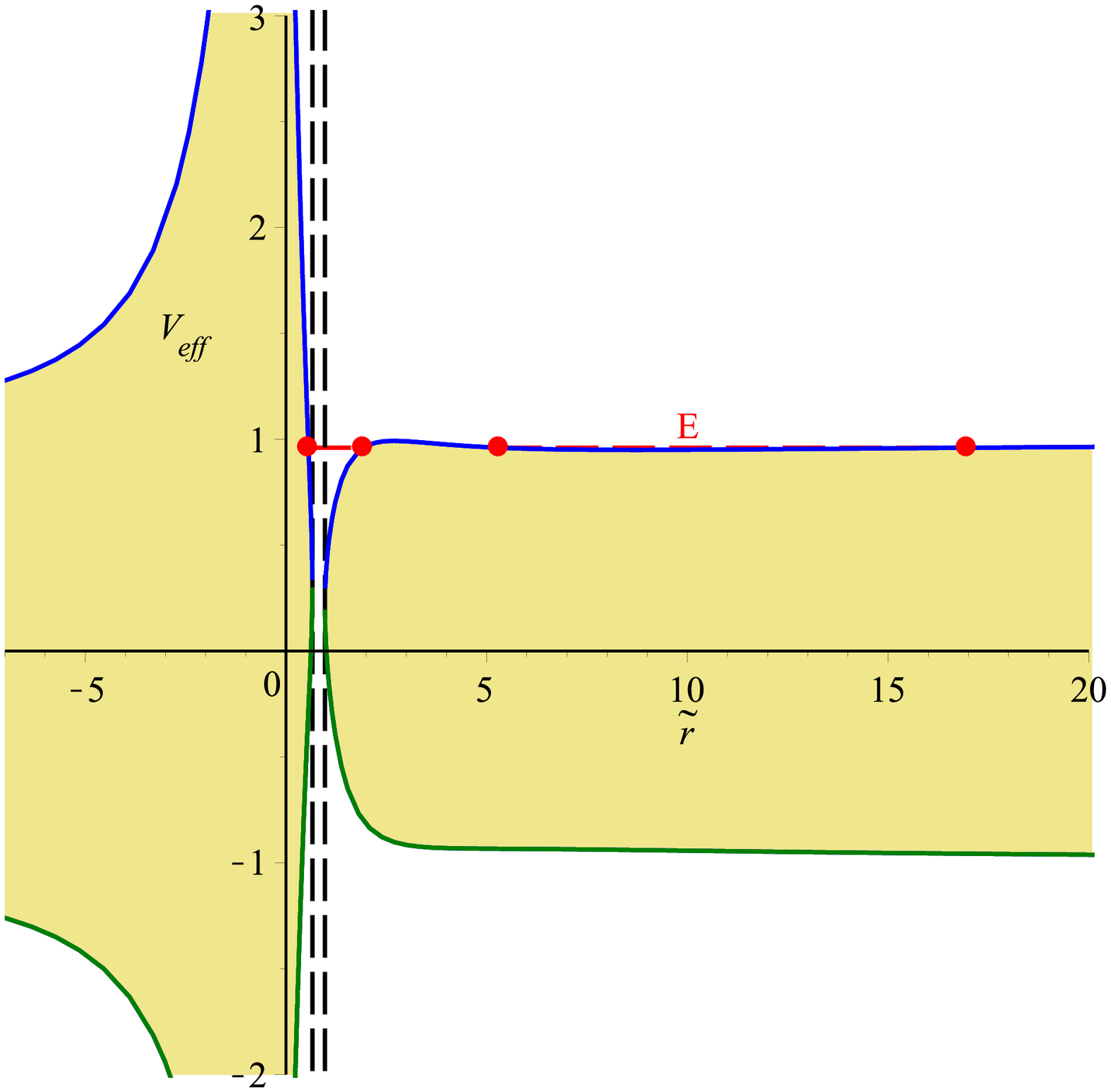}
	}
	\subfigure[closeup of figure (c)
]{
		\includegraphics[width=0.33\textwidth]{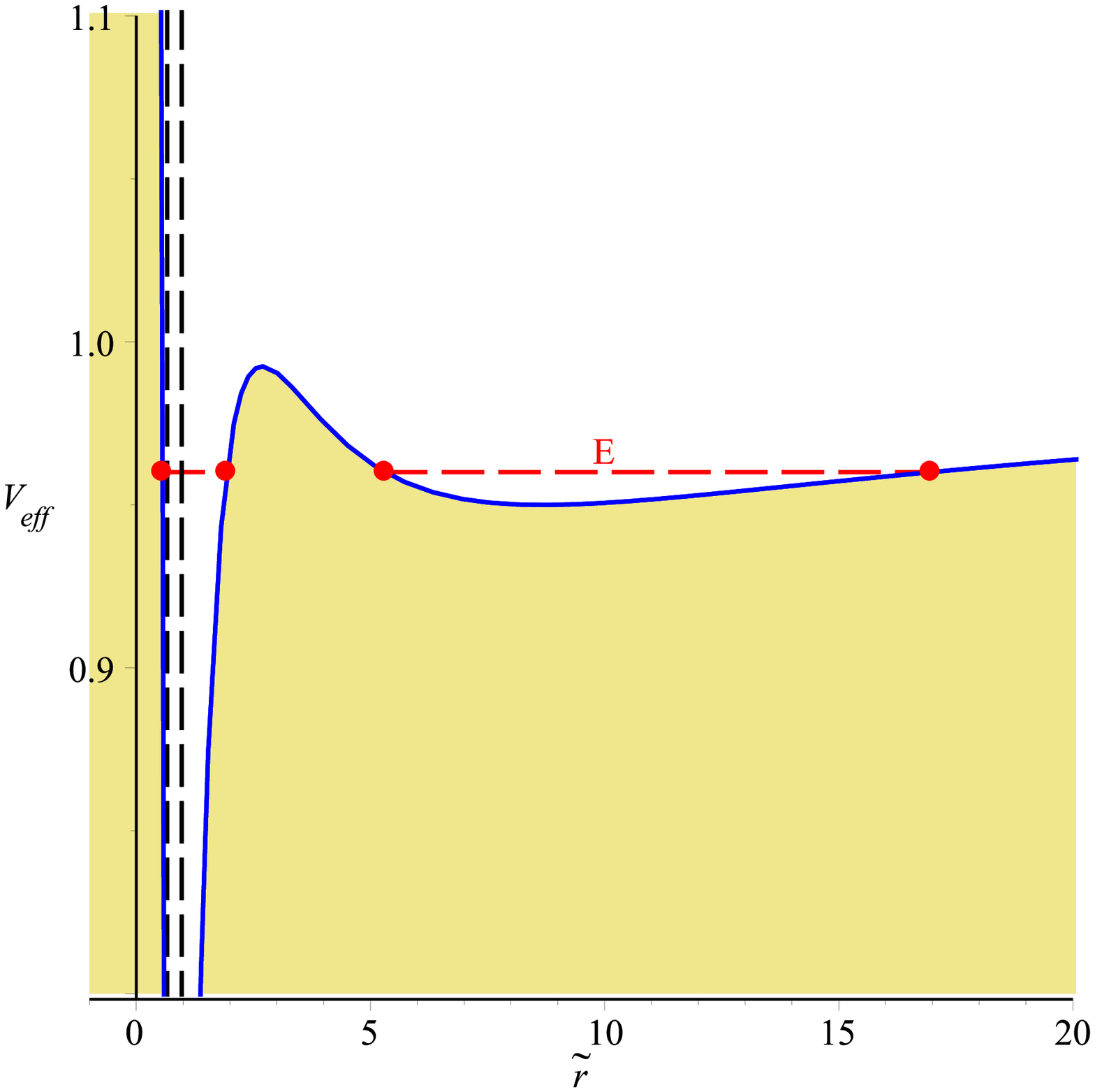}
	}
	\subfigure[]{
		\includegraphics[width=0.33\textwidth]{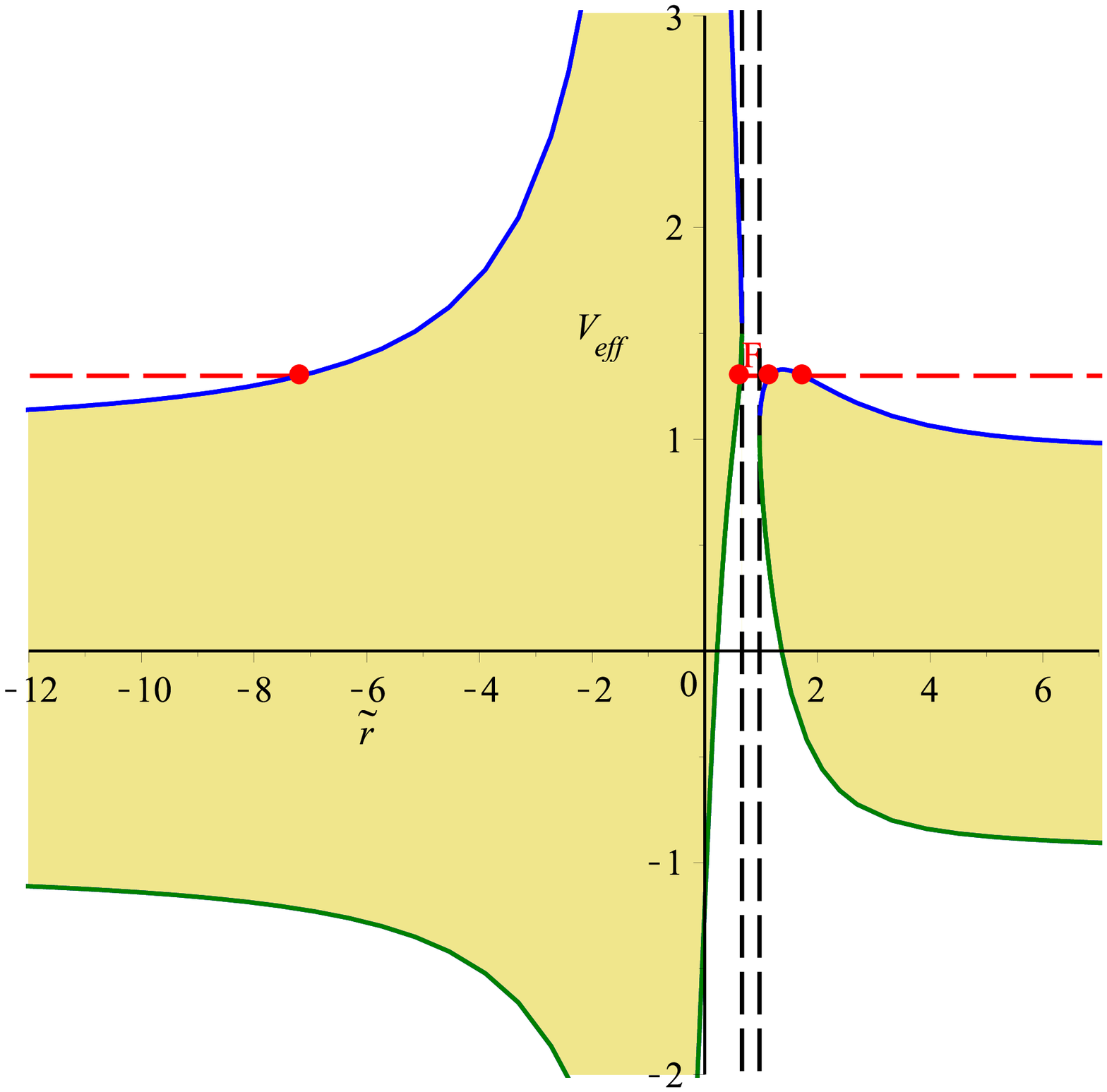}
	}
	\subfigure[closeup of figure (e)
]{
		\includegraphics[width=0.33\textwidth]{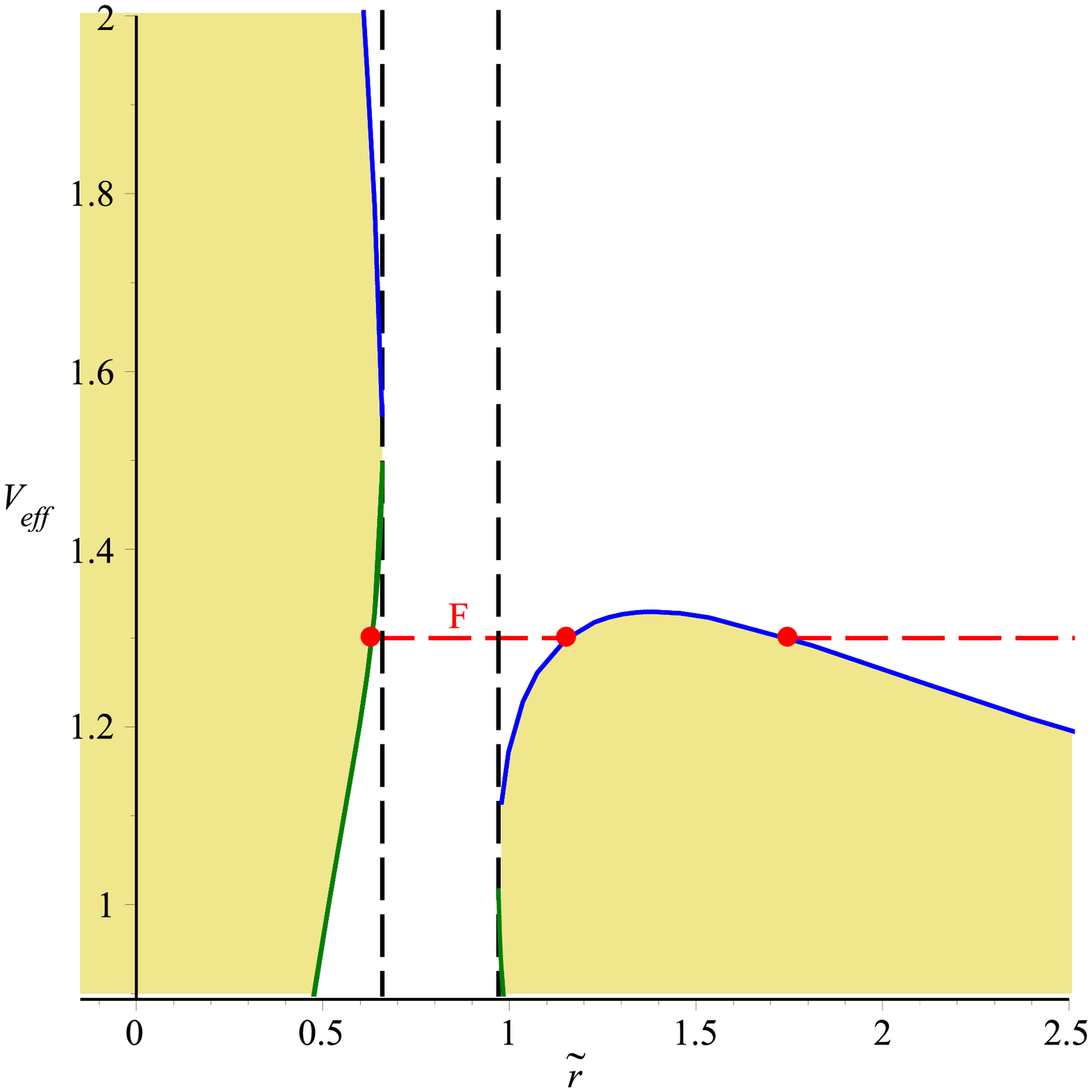}
	}
	\caption{ \label{pic:potential} Plots of the effective potential together with examples of energies for the different orbit types of table \ref{tab:orbit-types}. The blue and green curves represent to the two branches of the effective potential. The red dashed lines correspond to energies. The red dots mark the zeros of the polynomial $R$, which are the turning points of the orbits. In the khaki area no motion is possible since $\tilde{R} <0$. The vertical black dashed lines show the position of the horizons. $\epsilon=1$, $\tilde{a}=0.8$, $\tilde{K}=12$, $ \tilde{r}_{\alpha}=0.35 $, and $\tilde{L}=0.35$, $\tilde{L}=0.45$, $\tilde{L}=0.5$, $\tilde{L}=2.5$, for (a), (b), (c) and (e) respectively. }
\end{figure}

Combination of Figs.~\ref{fig:parameterplot-theta} and~\ref{pic:parametric-diagrams1} , is shown in Fig.~\ref{pic:parametric-diagrams}. For each regions, examples of effective potentials are demonstrated in Fig.~\ref{pic:potential}. With the help of the analytical solutions, the parametric $\tilde{L}$-$E^2$-diagrams and effective potentials, we can plot the orbits of test particles and light rays. 
Below we give a list of possible orbits. Let $\tilde{r}_+$ be the outer event horizon and $\tilde{r}_-$ be the inner horizon.
\begin{enumerate}
	\item \textit{Transit orbit} (TrO) with range $\tilde{r} \in (-\infty, \infty)$.
	\item \textit{Escape orbit} (EO) with range $\tilde{r} \in [r_1, \infty)$ with $r_1>\tilde{r}_+$, or with range $\tilde{r} \in (-\infty, r_1]$ with  $r_1<0$.
	\item \textit{Two-world escape orbit} (TEO) with range $[r_1, \infty)$ where $0<r_1 < r_-$.
	\item \textit{Crossover two-world escape orbit} (CTEO) with range $[r_1, \infty)$ where $r_1 < 0$.
	\item \textit{Bound orbit} (BO) with range $\tilde{r} \in [r_1, r_2]$ with
	\begin{enumerate}
		\item $r_1, r_2  > r_+$, or 
		\item $ 0 < r_1, r_2 < r_-$.
	\end{enumerate}
	\item \textit{Many-world bound orbit} (MBO) with range $\tilde{r} \in [r_1, r_2]$ where $0<r_1 \leq r_-$ and $r_2 \geq r_+$.
	\item \textit{Terminating orbit} (TO) with ranges either $\tilde{r} \in [0, \infty)$ or $\tilde{r} \in [0, r_1]$ with
	\begin{enumerate}
		\item $r_1\geq \tilde{r}_+$, or 
		\item $0<r_1<\tilde{r}_-$.
	\end{enumerate}
\end{enumerate}
It should be noted that, the only way for a geodesic to reach the singularity (Terminating Orbit) is $\tilde{R}(\tilde{r}=0)=0$ and $\tilde{\Theta}(\theta=\frac{\pi}{2})=0$. This is the case if $\tilde{K}=(E\tilde{a}-\tilde{L})^2$.\\
A summary of possible orbit types can be found in Table \ref{tab:orbit-types}.

\begin{figure}
 \centering
 \subfigure[]{
  \includegraphics[width=0.4\textwidth]{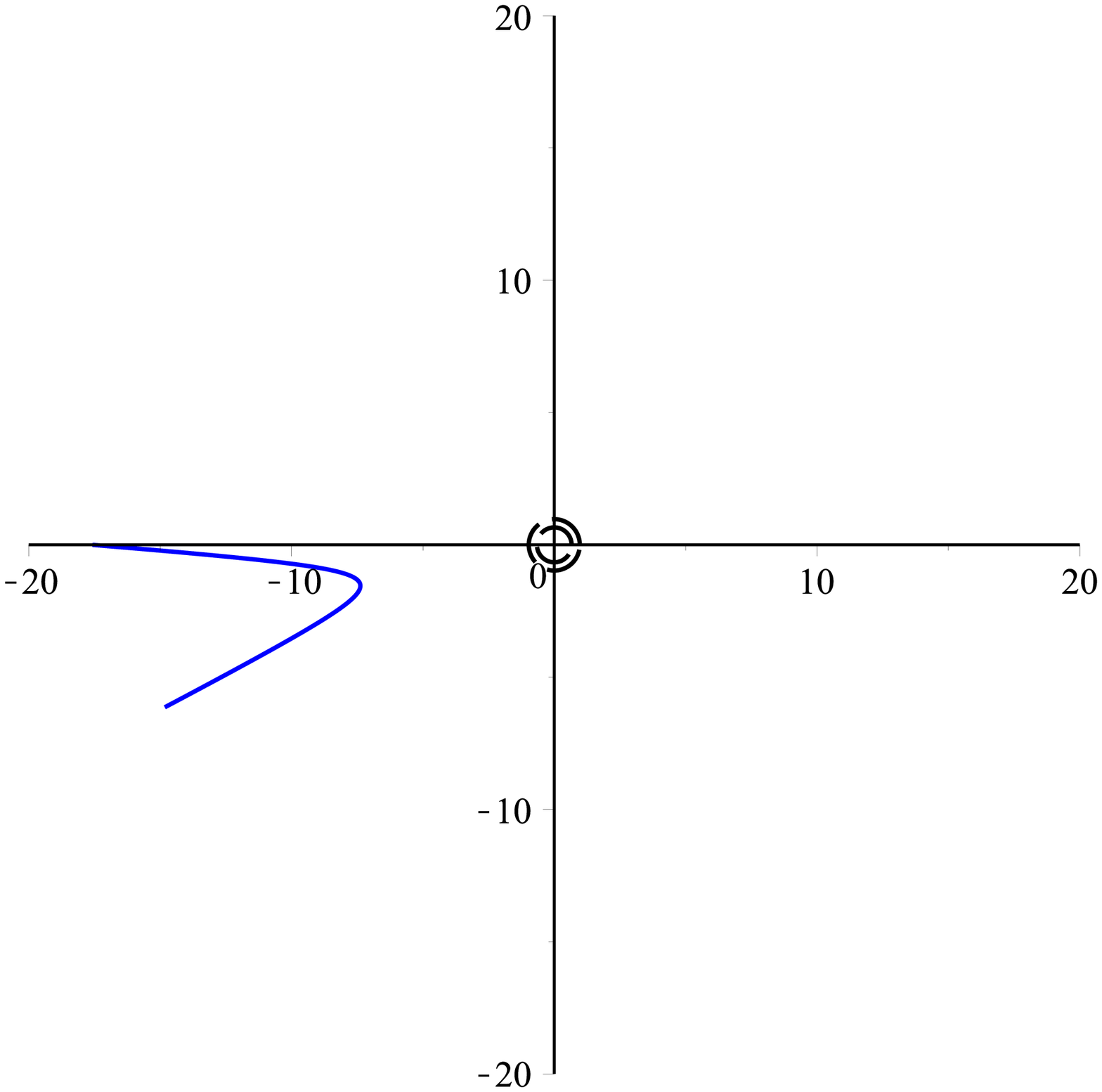}
 }
 \subfigure[]{
  \includegraphics[width=0.4\textwidth]{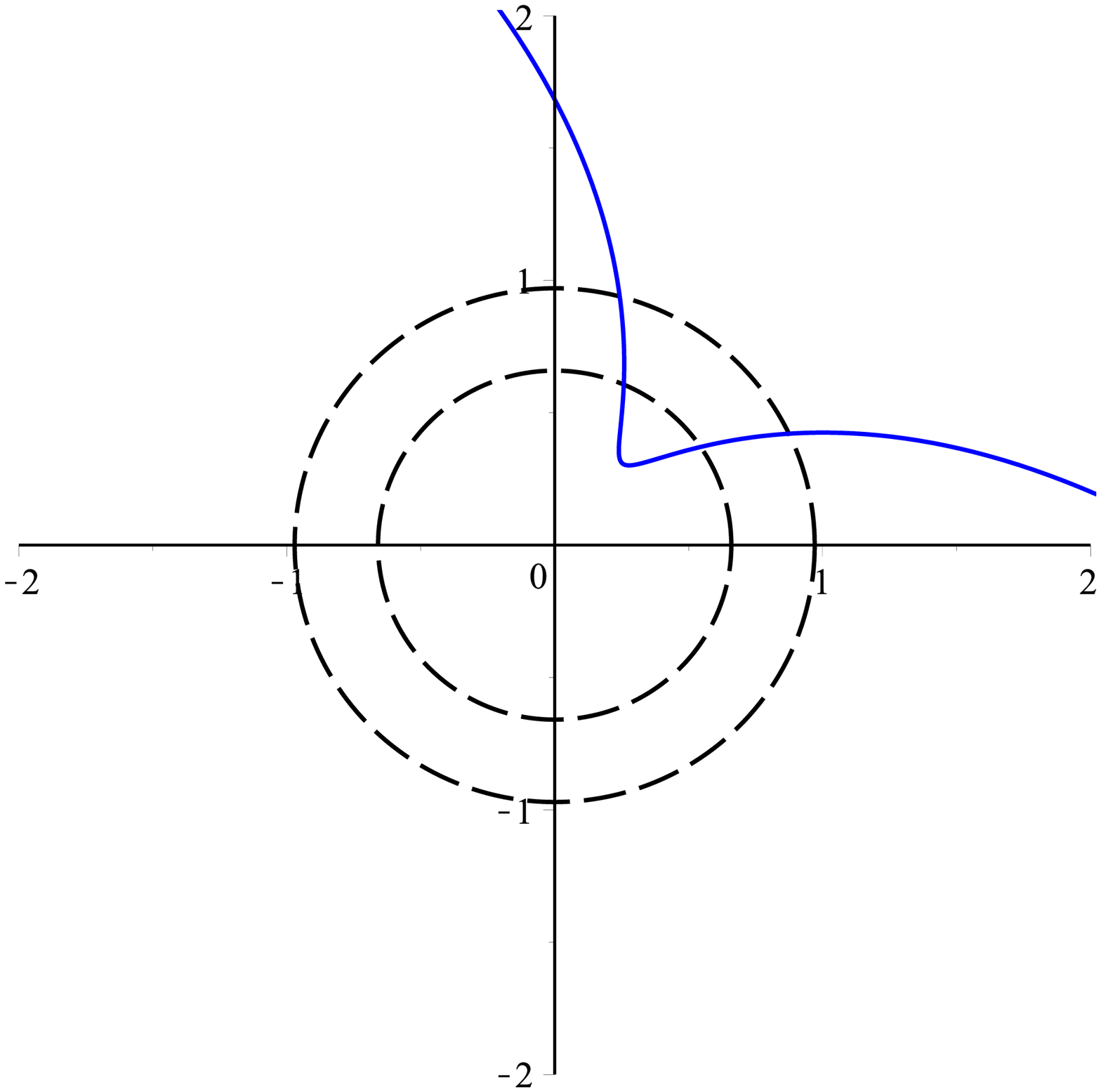}
 }
 \caption{Two examples of possible orbits in the spacetime of a rotating Kerr-Sen Dilaton-Axion black hole. An escape orbit (a) and a two-world escape orbit (b), with parameters $\epsilon=1$, $\tilde{a}=0.8$, $\tilde{K}=12$, $ \tilde{r}_{\alpha}=0.35 $, $L=0.35$, $E=2.5$. The blue lines show the path of the orbits and the circles represents the event horizons.}
\label{pic:bo-eo}
\end{figure}

\begin{figure}
 \centering
 \subfigure[]{
  \includegraphics[width=0.4\textwidth]{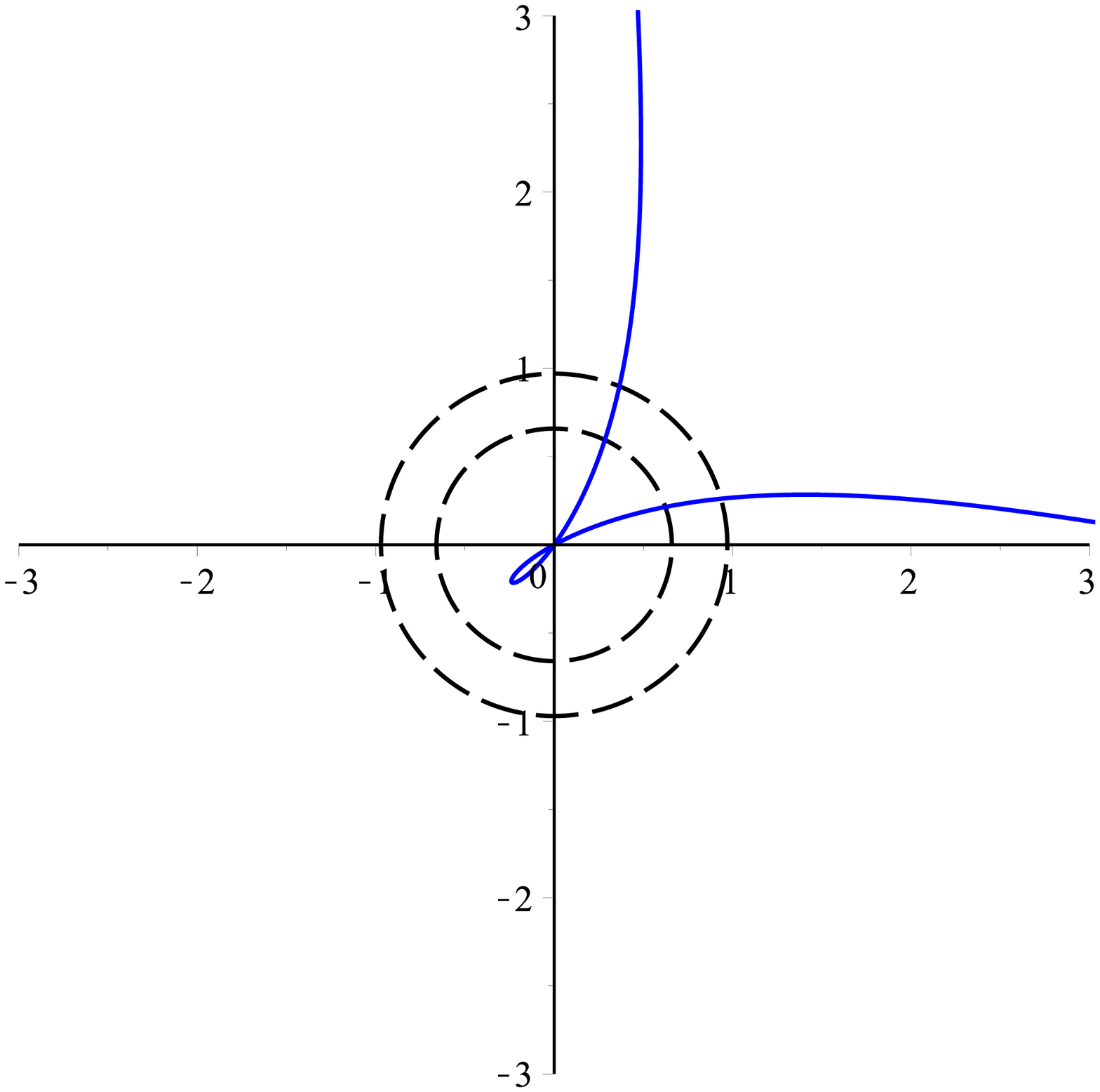}
 }
 \subfigure[]{
  \includegraphics[width=0.4\textwidth]{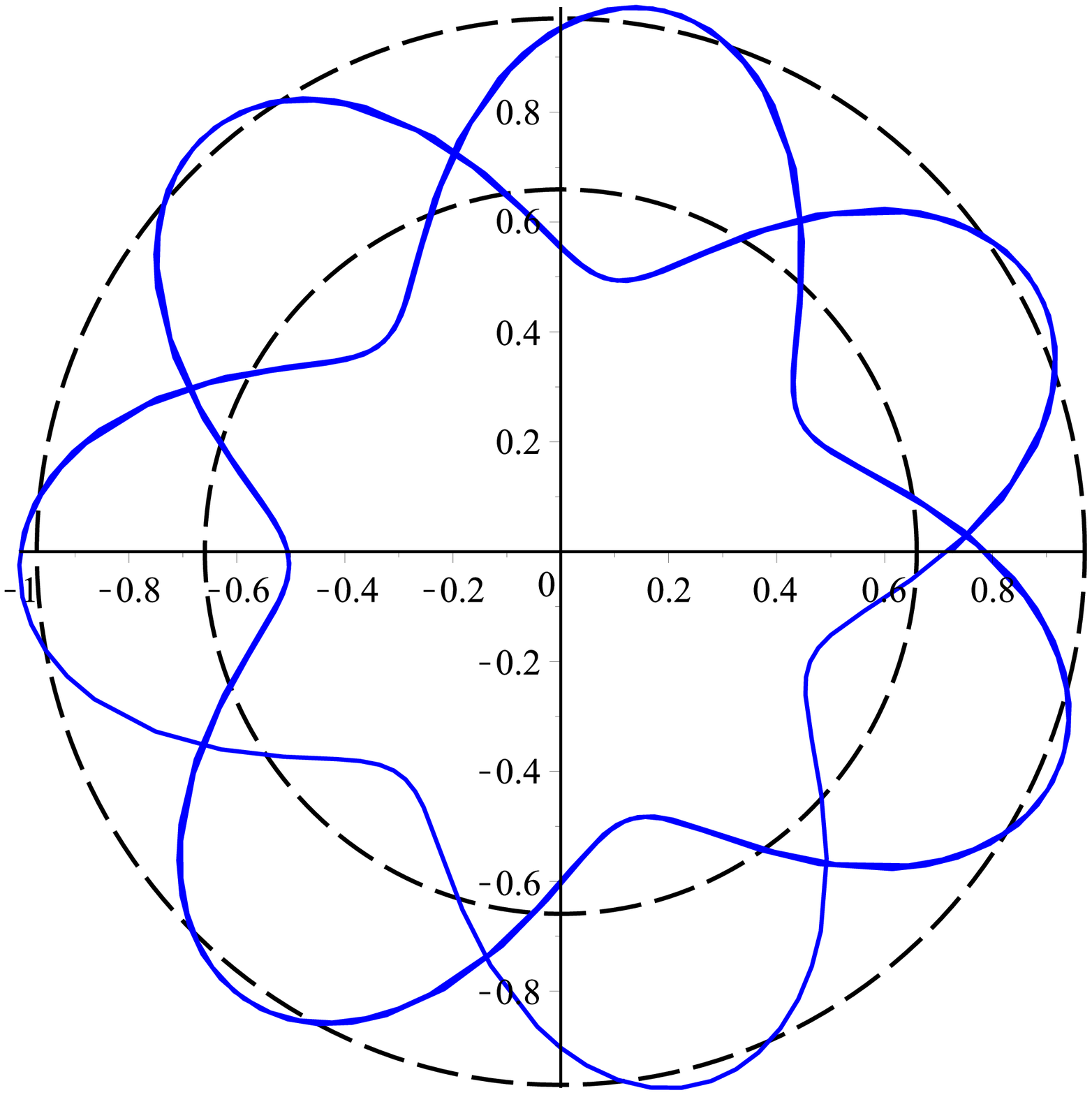}
 }
 \caption{Two examples of possible orbits in the spacetime of a rotating Kerr-Sen Dilaton-Axion black hole. A crossover two-world escape orbit (a) and a Mani-world escape orbit (b), with parameters $\epsilon=1$, $\tilde{a}=0.8$, $\tilde{K}=12$, $ \tilde{r}_{\alpha}=0.35 $, $L=0.35$, $E=7.5$ and $L=0.5$, $E=0.5$ respectively. The blue lines show the path of the orbits and the circles represent the inner and outer horizons.}
\label{pic:tro-teo}
\end{figure}

\begin{figure}
 \centering
 \subfigure[]{
  \includegraphics[width=0.4\textwidth]{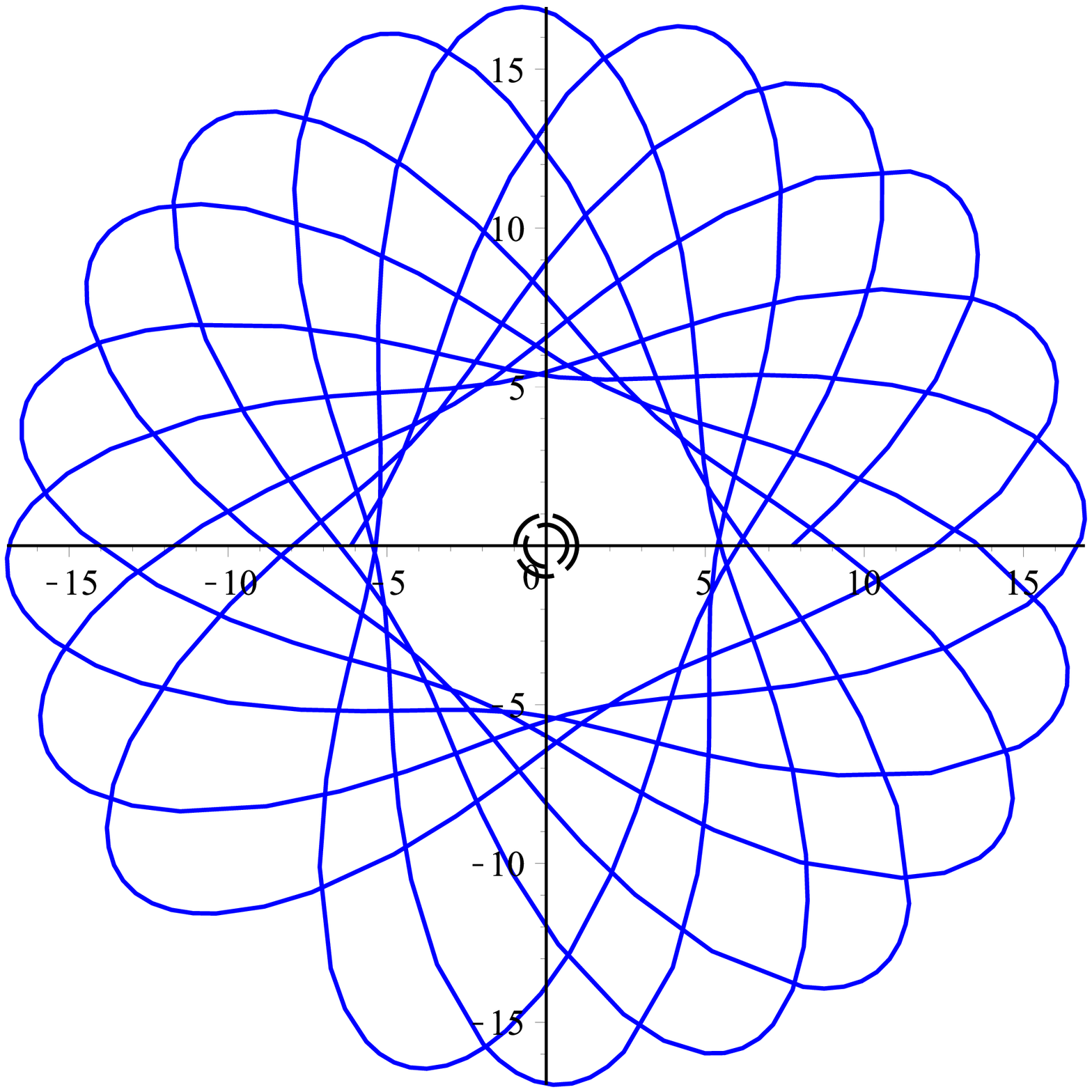}
 }
 \subfigure[]{
  \includegraphics[width=0.4\textwidth]{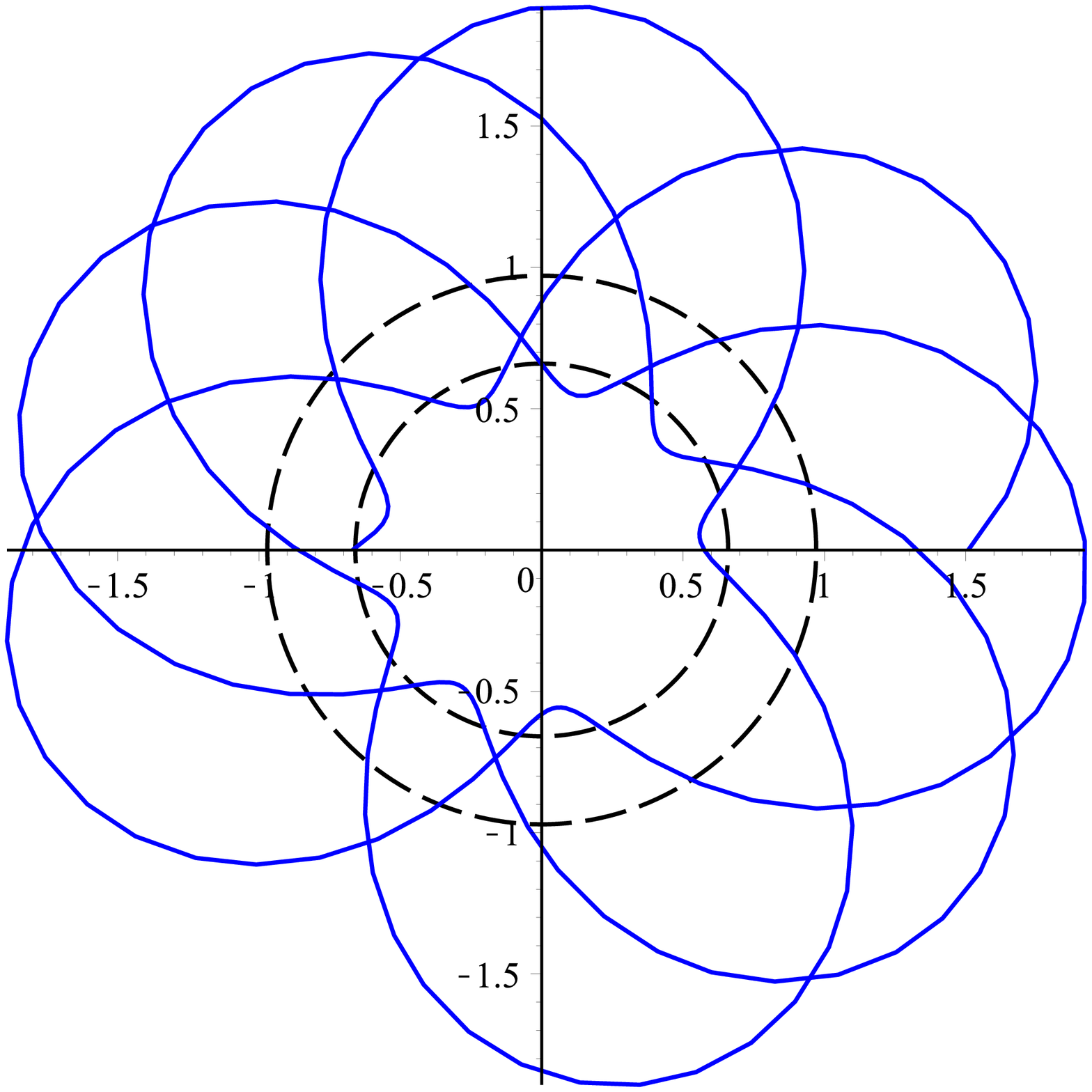}
 }
 \caption{Two examples of possible orbits in the spacetime of a rotating Kerr-Sen Dilaton-Axion black hole. A Bound orbit (a) and a Many-world bound orbit (b), with parameters $\epsilon=1$, $\tilde{a}=0.8$, $\tilde{K}=12$, $ \tilde{r}_{\alpha}=0.35 $, $L=0.5$, $E=0.96$.  The blue lines show the path of the orbits and the circles represent the inner and outer horizons.}
\label{pic:innerbo-mbo}
\end{figure}

Figure \ref{pic:bo-eo}, shows two example plots of an escape orbit (a) and a two-world escape orbit (b),  which crosses both horizons twice and escapes to another universe. A crossover two-world escape orbit, which crosses both horizons and $r=0$ and escapes to another universe, can be seen in Figure \ref{pic:tro-teo} (a). A Mani-world escape orbit which crosses both horizons several times, is depicted in \ref{pic:tro-teo} (b). In figure \ref{pic:innerbo-mbo} (a) a bound orbit outside the both horizons is shown. Figure \ref{pic:innerbo-mbo} (b) shows a many-world bound orbit, where both horizons are crossed several times.

%%%%%%%%%%%%
\begin{table}[h]
\begin{center}
\begin{tabular}{|lcccl|}\hline
type & zeros & region  & range of $\tilde{r}$ & orbit \\
\hline\hline
A & 0 & Ic &
\begin{pspicture}(-4,-0.2)(3.5,0.2)%\psgrid
\psline[linewidth=0.5pt]{->}(-4,0)(3.5,0)
\psline[linewidth=0.5pt](-2.5,-0.2)(-2.5,0.2)
\psline[linewidth=0.5pt,doubleline=true](-0.5,-0.2)(-0.5,0.2)
\psline[linewidth=0.5pt,doubleline=true](1,-0.2)(1,0.2)
\psline[linewidth=1.2pt]{-}(-4,0)(3.5,0)
\end{pspicture}
  & TrO
\\  \hline
B & 2 & IIc &
\begin{pspicture}(-4,-0.2)(3.5,0.2)%\psgrid
\psline[linewidth=0.5pt]{->}(-4,0)(3.5,0)
\psline[linewidth=0.5pt](-2.5,-0.2)(-2.5,0.2)
\psline[linewidth=0.5pt,doubleline=true](-0.5,-0.2)(-0.5,0.2)
\psline[linewidth=0.5pt,doubleline=true](1,-0.2)(1,0.2)
\psline[linewidth=1.2pt]{-*}(-4,0)(-3.5,0)
\psline[linewidth=1.2pt]{*-}(-3,0)(3.5,0)
\end{pspicture}
  & EO, CTEO
\\ \hline
C  & 2 & IIb &
\begin{pspicture}(-4,-0.2)(3.5,0.2)%\psgrid
\psline[linewidth=0.5pt]{->}(-4,0)(3.5,0)
\psline[linewidth=0.5pt](-2.5,-0.2)(-2.5,0.2)
\psline[linewidth=0.5pt,doubleline=true](-0.5,-0.2)(-0.5,0.2)
\psline[linewidth=0.5pt,doubleline=true](1,-0.2)(1,0.2)
\psline[linewidth=1.2pt]{-*}(-4,0)(-3,0)
\psline[linewidth=1.2pt]{*-}(-1.5,0)(3.5,0)
\end{pspicture}
& EO, TEO
\\ \hline
D & 2 & IIIb &
\begin{pspicture}(-4,-0.2)(3.5,0.2)%\psgrid
\psline[linewidth=0.5pt]{->}(-4,0)(3.5,0)
\psline[linewidth=0.5pt](-2.5,-0.2)(-2.5,0.2)
\psline[linewidth=0.5pt,doubleline=true](-0.5,-0.2)(-0.5,0.2)
\psline[linewidth=0.5pt,doubleline=true](1,-0.2)(1,0.2)
%\psline[linewidth=1.2pt]{-*}(-4,0)(-3,0)
\psline[linewidth=1.2pt]{*-*}(-0.5,0)(1.5,0)
%\psline[linewidth=1.2pt]{*-}(2,0)(3.5,0)
\end{pspicture}
  & TO/MBO
\\ \hline
E & 4 & IVb &
\begin{pspicture}(-4,-0.2)(3.5,0.2)%\psgrid
\psline[linewidth=0.5pt]{->}(-4,0)(3.5,0)
\psline[linewidth=0.5pt](-2.5,-0.2)(-2.5,0.2)
\psline[linewidth=0.5pt,doubleline=true](-0.5,-0.2)(-0.5,0.2)
\psline[linewidth=0.5pt,doubleline=true](1,-0.2)(1,0.2)
%\psline[linewidth=1.2pt]{-*}(-4,0)(-3,0)
%\psline[linewidth=1.2pt]{*-*}(-2,0)(-1.5,0)
\psline[linewidth=1.2pt]{*-*}(2,0)(3,0)
\psline[linewidth=1.2pt]{*-*}(-1,0)(1.5,0)
\end{pspicture}
  & MBO, BO
\\ \hline
F & 4 & Vb &
\begin{pspicture}(-4,-0.2)(3.5,0.2)%\psgrid
\psline[linewidth=0.5pt]{->}(-4,0)(3.5,0)
\psline[linewidth=0.5pt](-2.5,-0.2)(-2.5,0.2)
\psline[linewidth=0.5pt,doubleline=true](-0.5,-0.2)(-0.5,0.2)
\psline[linewidth=0.5pt,doubleline=true](1,-0.2)(1,0.2)
\psline[linewidth=1.2pt]{-*}(-4,0)(-3,0)
%\psline[linewidth=1.2pt]{*-*}(-2,0)(-1.5,0)
\psline[linewidth=1.2pt]{*-*}(-1,0)(1.5,0)
\psline[linewidth=1.2pt]{*-}(2,0)(3.5,0)
\end{pspicture}
& EO, MBO, EO
\\ \hline\hline
\end{tabular}
\caption{\label{tab:orbit-types} Types of orbits in the spacetime of a rotating Kerr-Sen Dilaton-Axion black hole. The range of the orbits is represented by thick lines. The dots show the turning points of the orbits. The positions of the horizons are marked by a vertical double line. The single vertical line indicates $\tilde{r}=0$.}
\end{center}
\end{table}
%%%%%%%%%%%%%%%%%%%%%%

\clearpage

%%%%%%%%%%%%%%%

\section{CONCLUSIONS}\label{conclusions}

In this paper, we considered the motion of test particles and light rays 
in the spacetime of the static (GMGHS, magnetically charged GMGHS and
electrically charged GMGHS) and the rotating (Ker-Sen Dilaton-Axion
) dilaton black holes. We have derived geodesic equations of motion and classified them
according to their energy $E$ and angular momentum $L$. The geodesic
equations of motion can be solved in terms of the elliptic
Weierstrass $ \wp $, $\zeta$ and $\sigma$ functions. 
Possible types of orbits were derived using analytical solutions, effective potential
techniques and parametric diagrams. For electrically charged GMGHS
black hole, EO, TO and BO are possible, while for GMGHS and
magnetically charged GMGHS black holes, EO, TEO, BO and MBO are 
possible and any type of terminating orbit are not possible for these metrics.
Also, for rotating (Ker-Sen Dilaton-Axion) dilaton black hole, TrO, 
EO, TEO, CTEO, TO, BO and MBO are possible.
Some observational phenomena such as the periastron shift 
of bound orbits and the deflection angle  of light, are some results of these 
solutions. At the end of sec.~\ref{Static Dilaton}, we calculated 
such Astrophysical applications. Moreover, it would be interesting to use 
the results of this paper to study the shadow of dilaton black holes.

%%%%%%%%%%%%%

\begin{acknowledgements}

We would like to thank anonymous referee for useful comments. 
Also we would like to thank Bahareh Hoseini for helpful discussions 
and her guidance.

\end{acknowledgements}

%%%%%%%%%%

\bibliographystyle{amsplain}
%%%%%%%%%%%%%%%%%%%%%%%%%%%%%%%%%%%%%%%%%%%

\end{document}